\documentclass[notitlepage,twocolumn,superscriptaddress]{revtex4-2}
\usepackage[utf8]{inputenc}
\usepackage{graphicx}
\usepackage{url}
\usepackage[hidelinks]{hyperref} 

\begin{document}

\title[]{The Martini Model in Materials Science}
\author{Riccardo Alessandri}
\affiliation{Zernike Institute for Advanced Materials and 
Groningen Biomolecular Sciences and Biotechnology Institute,
University of Groningen, Nijenborgh 7, 9747 AG Groningen, The Netherlands}
\affiliation{These two authors contributed equally}
\author{Fabian Gr{\"u}newald}
\affiliation{Zernike Institute for Advanced Materials and 
Groningen Biomolecular Sciences and Biotechnology Institute,
University of Groningen, Nijenborgh 7, 9747 AG Groningen, The Netherlands}
\affiliation{These two authors contributed equally}
\author{Siewert J. Marrink}
\email{s.j.marrink@rug.nl}
\affiliation{Zernike Institute for Advanced Materials and 
Groningen Biomolecular Sciences and Biotechnology Institute,
University of Groningen, Nijenborgh 7, 9747 AG Groningen, The Netherlands}

\begin{abstract}
The Martini model, a coarse-grained force field initially developed with biomolecular 
simulations in mind, has found an increasing number of applications in the
field of soft materials science. 
The model's underlying building block principle does not pose
restrictions on its application beyond biomolecular systems.
Here we highlight the main applications to date of the Martini model in materials
science, and we give a perspective for the future developments in this field, 
in particular in light of recent developments such as 
the new version of the model, Martini 3.
\end{abstract}

\maketitle


\section{\large Introduction}
Coarse-grained (CG) force fields have gained a lot of popularity 
in the field of molecular dynamics (MD) simulations~\cite{2014HIIngolfsson-WIREsComputMolSci, 
2015TakadaAccChemRes, 2020JoshiMolSimul}.
By averaging out some of the atomistic degrees of freedom, 
they allow for a substantial alleviation of both the spatial 
and temporal limitations of all-atom models.
The Martini model~\cite{2007SJMarrink-JPhysChemB, 2004SJMarrink-JPhysChemB, 
2013SJMarrink-ChemSocRev} is an example of a popular CG force field 
that has been incorporated by the worldwide user community to study 
a large variety of (bio)molecular processes~\cite{2013SJMarrink-ChemSocRev, 
2019Bruininks2019MIMB}.

In the Martini model, typically four heavy atoms with their associated 
hydrogen atoms are grouped into one functional group, denoted as a CG bead. 
This effectively reduces the number of particles to be simulated in a system, 
increasing the simulation speed. In addition, the smoother CG energy 
landscape leads to faster overall dynamics and allows the use of larger 
time steps compared to all-atom simulations. Together, this results in 
a significant increase in accessible length and time
scales of a few orders of magnitude, albeit at somewhat reduced level of accuracy.

The CG beads represent small chemical fragments, and are used as building 
blocks for larger molecules. Parametrization of the non-bonded interactions 
between the CG beads is based on reproducing thermodynamic data such 
asfree energies of transfer of organic compounds. In addition, reference 
all-atom simulation data are used to derive effective bonded terms. 
Such a combination of top-down and bottom-up approaches enables the 
Martini model to distinguish different chemical species and form a useful 
bridge between atomistic and macroscopic scales.

From the first applications, purely concerned with
lipids~\cite{2003SJMarrink-JAmChemSoc-self-assembly, 2003SJMarrink-JAmChemSoc-fusogenicity}, 
the Martini model has been applied to a vast amount of biomolecular systems. 
The compatibility with a wide library of existing molecules, 
which includes all major biomolecules such as
proteins~\cite{2008LMonticelli-JChemTheoryComput},
sugars~\cite{2009CALopez-JChemTheoryComput}, DNA~\cite{2015JJUusitalo-JChemTheoryComput}, 
or RNA~\cite{2017JJUusitalo-BiophysJ}, 
as well as an increasing amount of synthetic
molecules~\cite{2013SJMarrink-ChemSocRev}, is one of the key strengths 
of the Martini model. It enables researchers to easily simulate complex 
many-component systems and focus on more advanced simulation methodologies.

\begin{figure}[htbp]
  \centering
  \includegraphics[width=0.48\textwidth]{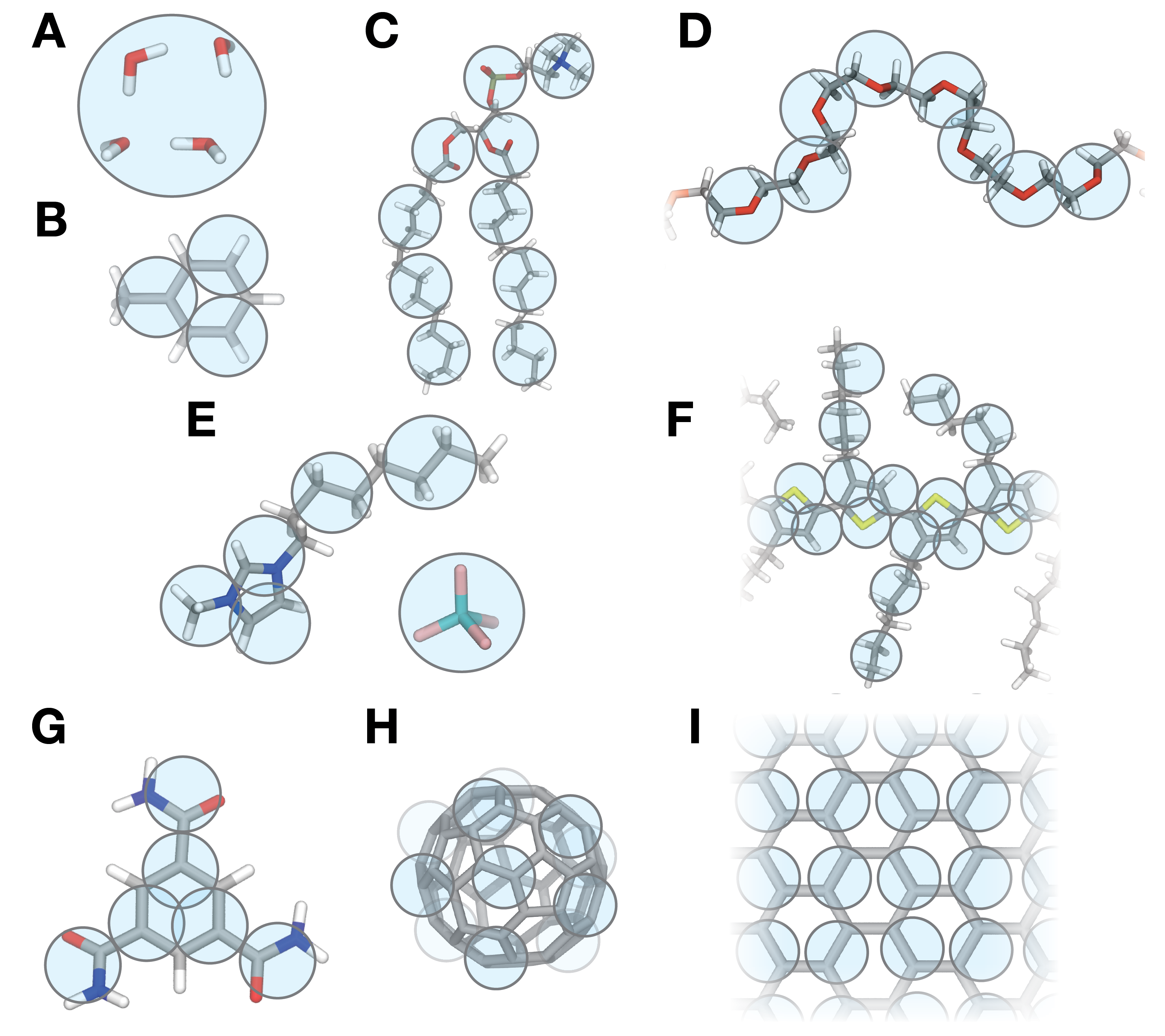} 
  \caption{Martini mapping examples of selected molecules. 
           (A) Standard water particle representing four water molecules; 
           (B) the organic solvent toluene;
           (C) dimyristoylphosphatidylcholine (DMPC) lipid;
           (D) polyethylene oxide (PEO); 
           (E) the 1-octyl-3-methylimidazolium tetrafluoroborate [C8Imim]$^+$[BF$_4$]$^-$ ionic liquid
           (F) poly(3-hexylthiophene) (P3HT);
           (G) the 1,3,5-benzenetricarboxamide (BTA) self-assembling molecule;
           (H) C$_{60}$ fullerene;
           (I) a graphene surface.
           Martini CG beads are shown as cyan transparent beads
           overlaying the atomistic structure.
  }
  \label{fig:mapping-materials}
\end{figure}

In recent years, the Martini model has found more and more 
applications in the field of materials science. 
In light of the underlying building block principle of the Martini model, 
there is no reason to restrict its applications to biomolecular systems.
Indeed already more than a decade ago the Martini model has been 
applied to simulate polymeric systems~\cite{2009HLee-JPhysChemB}.
In principle, any molecule can be represented by Martini CG beads, 
as illustrated in Figure~\ref{fig:mapping-materials}. 
Based on this versatility, the Martini model is nowadays used to 
simulate a wide range of materials, including 
(block co)polymers, nanoparticle-polymer composites,
organic electronic materials, ion-conducting materials,
self-assembled supramolecular materials, ionic liquids, and
others.

The use of CG models in material science is of course not new. In fact, 
some of the very first applications of CG models, such as the freely 
jointed chain models of Binder and coworkers \cite{1981BaumgartnerJCP}, 
already targeted polymer dynamics. Since those early days, a large plethora 
of CG models have been developed to model an equally large variety of 
materials\cite{2014PotestioEntropy, 2016KaratrantosPolymRev, 
2017GooneiePolymers, 2017MOSteinhauser-AIMSMaterSci, 2019ChenCurrOpChem, 
2019CasaliniGels, 2020Muller, 2020BinderJPhysMaterials}. 
Two major assets of the Martini model, that set it apart from most 
other CG models, are (i) the retaining of near atomic resolution, 
as opposed to more generic models that are frequently used to capture 
global system properties but are unable to offer a direct connection 
to chemical specificity, and (ii) the broad range of compatible parameters 
available for different classes of molecules, enabling simulations of 
the ever expanding group of complex and hybrid materials as well as 
the interaction of materials with biological systems.

In the following sections, after summarizing Martini parametrization strategies
tailored to material systems, we discuss the main application areas to date 
of the Martini model in materials science, illustrated with selected examples. 
While there are numerous Martini applications which involve the interaction
of synthetic materials with biomolecules, we will not cover those here.
The interested reader is referred to recent reviews and references 
therein~\cite{2013SJMarrink-ChemSocRev,2016GRossi-AdvPhysX,
2019TCasalini-FrontBioengBiotechnol}.
We conclude with a perspective for the future developments in this field, 
in particular in light of the new version of the model, Martini 3, as well as 
recently developed tools to generate starting structures for polymeric systems 
and to allow constant pH simulations.

\section{\large Parameterization strategies}
\subsection{\normalsize General guidelines}
Martini typically gathers groups of four non-hydrogen atoms in CG beads---see
Figure~\ref{fig:mapping-materials} for some representative mappings. 
The interactions between beads---described by a 12-6 Lennard-Jones (LJ) potential---represent 
the nature of the underlying chemical groups and have been systematically parametrized 
to reproduce free energies of transfer of solutes between polar and non-polar solvents. 
There are four main particle types: polar (P), intermediately polar (N), 
apolar (C) and charged (Q).
These types are in turn divided in subtypes based on their hydrogen-bonding 
capabilities (with a letter denoting: d = donor, a = acceptor, da = both, 0 = none) 
or their degree of polarity (with a number from 1 = low polarity to 5 = high polarity). 
This gives a total of 18 particle types: the Martini building blocks. 
Such a building block approach---a discrete number of particles which
interact using a limited number of interaction levels---provides compatibility
of different Martini models and facilitates parameterization of new molecules, 
albeit limiting the quantitative accuracy of the force field.
In addition to the regular Martini beads, smaller bead sizes (small and tiny beads) 
are used for groups that are represented at higher resolution such as ring-like 
fragments.~\cite{2007SJMarrink-JPhysChemB, 2013SJMarrink-ChemSocRev}.

In general, a Martini model for an arbitrary molecule can be generated as follows.
\textit{(1)} The atomistic structure is partitioned into a number of beads, 
maximizing the four-to-one mapping while preserving the symmetry of the molecule; 
smaller beads may be used to better represent the geometry of small ring-like fragments.
 \textit{(2)} Bead types describing the nonbonded interactions
of the models are determined by comparing to already existing fragments or by 
computing free energies of transfer and selecting the best matching bead. 
\textit{(3)} Bonded interactions, defined by a standard set of potential energy 
functions typical of classical force fields, are parametrized by comparing to
atomistic simulations or experimental data.
The basic assumption underlying the Martini approach is that the thoroughly parameterized properties 
of the individual Martini bead types are transferable to the molecule as a whole when linked
together to reproduce the overall topology of the desired molecule. 
This basic assumption entails some pitfalls~\cite{2019RAlessandri-JChemTheoryComput},
and hence requires validation, which commonly comes from either comparison to higher resolution
atomistic simulations or to experimental data~\cite{2013SJMarrink-ChemSocRev}.

Below we describe strategies based on the outlined general procedure but which 
have been found helpful for specific classes of molecules relevant to material systems 
such as polymers, nanoparticles, and surfaces. 
These may include specific reference data coming from atomistic simulations or experimental
measurements which have been found especially helpful to validate Martini models in this area.

\subsection{\normalsize Polymers}
Martini polymers are applied in a wide range of studies on biomolecular and 
materials science systems. As first suggested by \citeauthor{2012GRossi-Macromolecules-PS}, 
parameters for Martini polymers are ideally generated by matching: 
\textit{(1)} the free energy of transfer of the monomeric repeat unit; 
\textit{(2)} bond and angle distributions of atomistic reference simulations; and
\textit{(3)} long-range structural properties.~\cite{2012GRossi-Macromolecules-PS} 
Overall many carefully parametrized Martini polymers, such as 
polystyrene (PS)~\cite{2011GRossi-SoftMatter-PS}, PEO~\cite{2018FGrunewald-JPhysChemB}, 
polyethylene (PE)~\cite{2015EPanizon-JPhysChemB}, and polypropylene (PP)~\cite{2015EPanizon-JPhysChemB}, 
have been validated by showing that they are able to reproduce a number of single-chain properties. 
For example, the PS, PEO, as well as PEO-PPO models reproduce structural properties 
such as radii of gyration in different solvents~\cite{2011GRossi-SoftMatter-PS, 
2018FGrunewald-JPhysChemB, 2014SNawaz-JPhysChemB}.
Indeed, validation of these properties in multiple solvents---so as to probe
good, bad, and theta solvent conditions---is desirable.
Furthermore, persistence lengths, structure factors, and polymer melt density 
consistute other properties which may be used as validation targets.

An important aspect to keep in mind when parametrizing (Martini) CG polymer models is that
the use of torsion angle potentials borrowed from atomistic MD simulations is often 
unsuitable~\cite{2013MBulacu-JChemTheoryComput}. 
Given the softer nature of angle potentials in CG models, conformations
which lead to numerical instabilities can be encountered much more often, leading to
impractical simulations. \citeauthor{2013MBulacu-JChemTheoryComput} have devised strategies,
like the use of special bonded potentials such as the Restricted
Bending Potential (ReB)---implemented in GROMACS~\cite{2015MJAbraham-GMX5}---or 
virtual site-aided definition of bonded terms, 
to combat this problem~\cite{2013MBulacu-JChemTheoryComput}.
These strategies allow to avoid such instabilities and have been 
successfully applied to PEO~\cite{2013MBulacu-JChemTheoryComput,2018FGrunewald-JPhysChemB}, 
PE~\cite{2015EPanizon-JPhysChemB} and other models. Using these strategies,
simulations with chains of 500 repeat units over several 10s of microseconds can easily
be realized~\cite{2018FGrunewald-JPhysChemB}.

The current library of Martini polymers comprises more than 50
different polymers.
The models available range from simple linear polymers such as
PEO (Figure~\ref{fig:mapping-materials}D)~\cite{2009HLee-JPhysChemB, 2012GRossi-JPhysChemB-PEG, 
2014EChoi-JPhysChemB, 2017TTaddese-JPhysChemB, 2018FGrunewald-JPhysChemB}
and PS~\cite{2011GRossi-SoftMatter-PS, 2012GRossi-Macromolecules-PS,
2019YMBeltukov-EurPhysJD, 2019JHHung-SoftMatter, 2020SHMin-JPhysChemB}, 
over branched and hyper-branched polymers 
such as (grafted) polyamidoamine (PAMAM)~\cite{2006HLee-JPhysChemB-PAMAM, 2008HLee-JPhysChemB-PAMAM,
2011HLee-Macromolecules, 2011HLee-Macromolecules2, 2011TZhong-FluidPhEquilibria, 2011WTian-SoftMatter} 
or polyethylenimine (PEI) dendrimers~\cite{2017HJeong-SciRep, 
2019TABeu-ChemPhysLett,2019SMahajan-JComputChem},
to conjugated polymers such as poly(3-hexylthiophene) 
(P3HT, Figure~\ref{fig:mapping-materials}F)~\cite{2017RAlessandri-JAmChemSoc}
and block copolymers~\cite{2014SNawaz-JPhysChemB,2014JCJohnson-JPhysChemB,
2019GCampos-Villalobos-MolSystDesEng} such as PEO-PPO~\cite{2014SNawaz-JPhysChemB}. 
Moreover, polymers have also been developed within the Dry 
Martini~\cite{2015CArnarez-JChemTheoryComput} version of the force field: examples 
include PEO~\cite{2015SWang-Macromolecules}, PAMAM dendrimers~\cite{2016LChong-JComputChem},
and charged polysaccharide~\cite{2017CWen-ACSOmega}.
The fine degree of coarse-graining of the Martini model means that there is 
no limitation to the topology of the polymer.
Recently, a tool for easily generating topologies, single chain starting structures,
and morphologies of polymers has been developed by Gr{\"u}newald and 
co-workers---Polyply~\cite{SOONFGrunewald-Polyply, 2021FGrunewald-chapter}. 
Polyply's internal library already contains several Martini polymer models from the 
literature but more models can be contributed via GitHub~\cite{SOONFGrunewald-Polyply}.
This tool standardizes and greatly simplifies the generation and setup of
Martini polymer systems, as discussed more extensively in the Outlook section.

\subsection{\normalsize Nanoparticles}
Martini models for fullerene~\cite{2008JWong-Ekkabut-NatNanotechnol,
2012LMonticelli-JChemTheoryComput, 2009RSGDOrzario-Nanoscale}, carbon 
nanotubes (CNTs)~\cite{2007EJWallace-NanoLett, 2011NPatra-JAmChemSoc,
2012HLee-JPhysChemC, 2013MLelimousin-Small},
graphene~\cite{2010AVTitov-ACSNano,2019MShi-JApplPhys,2020CDWilliams-2DMater} or 
MXene~\cite{2019JLi-AdvMaterInterfaces} flakes, clay nanoparticles~\cite{2019PKhan-JPhysChemB},
and functionalized nanoparticles~\cite{2011J-QLin-JPhysChemC, 2013JDong-MacromolTheorySimul,
2020MPannuzzo-NanoLett} 
have also been developed to study their interaction with
both other synthetic or biomolecular systems.
The C$_{60}$ fullerene model developed by \citeauthor{2012LMonticelli-JChemTheoryComput}
represents a good parametrization strategy for nanoparticles.
The model has been developed by matching experimental free energies of transfer
between a wide range of solvents of different polarity and potentials of mean force (PMF) 
of dimerization in water and octane~\cite{2012LMonticelli-JChemTheoryComput}. This thorough
parametrization allowed the Martini fullerene model to reasonably reproduce 
solid-state properties and lead to translocation PMF across a lipid membrane
in good agreement with atomistic reference data. 
We note that the final model did not use a standard Martini CG bead but instead 
required some refinements. The thorough refinement of the parameters across a wide range
of solvents, however, resulted in a model which could be used in other solvents,
such as chlorobenzene, where a comparison to atomistic reference data showed excellent 
agreement~\cite{2017RAlessandri-JAmChemSoc} even though the C60 model had not been 
tested explicitly in chlorobenzene at the time of development. 
Hence, validation of nanoparticle models in different solvents, by means of comparison
to experimental transfer free energies when available and PMFs of dimerization in different
solvents is desirable when developing Martini models for these systems.

\subsection{\normalsize Surfaces}
Simulations of Martini systems in materials science have also lead
to the development of models for graphite~\cite{2012DSergi-JChemPhys, 
2013CGobbo-JPhysChemC}, graphene~\cite{2012DWu-JPhysChemB},
and silica~\cite{2018EPerrin-JPhysChemB}.
As an example, the parametrization of the graphite model by \citeauthor{2013CGobbo-JPhysChemC}
used as reference experimental data enthalpies due to lack of experimental free
energy data~\cite{2013CGobbo-JPhysChemC}. Namely, the authors used
(1) enthalpies of adsorption of individual molecules from the gas phase 
on graphite, (2) wetting enthalpies of pure liquids, and (3) enthalpies 
of displacement of solutes (long-chain organic molecules) from different solvents
(heptane and phenyloctane) to graphite.
The model required the development of a custom bead representing graphite
with a 2-to-1 mapping scheme, which eventually could achieve 
semi-quantitative reproduction of the experimental enthalpies~\cite{2013CGobbo-JPhysChemC}.

Besides the parametrization strategies and validation targets outlined
above, several other application-specific critical tests can be carried out to
validate a particular Martini model. As we describe the applications in materials
science to date in the following sections, we invite the reader to check
the reference of interest to find out about further validation targets for the
application of interest.

\section{\large Example applications}
\subsection{\normalsize Polymeric hydrogels}
Polymeric hydrogels are networks composed of hydrophilic polymers that are  
covalently or physically cross-linked. These polymer networks can swell 
taking up a multiple of their dry weight in water~\cite{2012ASHoffman-AdvDrugDelivRev}. 
They are frequently employed in drug delivery either as nanogel or macroscopic 
material~\cite{2012ASHoffman-AdvDrugDelivRev,Mauri2018}. 
Martini simulations were employed to understand: (1) interactions of hydrogels 
with the cargo molecules at molecular detail~\cite{Zadok2018} (2) the gel response 
to environmental effects such as pH~\cite{2017HXu-ChemCommun}; (3) transport properties 
of the cargo inside a gel~\cite{2014HSalahshoor-JMechBehavBiomed}, and 
(4) effect of the salt concentration on the gel~\cite{2020RKumar-IntJMolSci,2020KYue-MolSimul}.

\begin{figure}[htbp]
  \centering
  \includegraphics[width=0.48\textwidth]{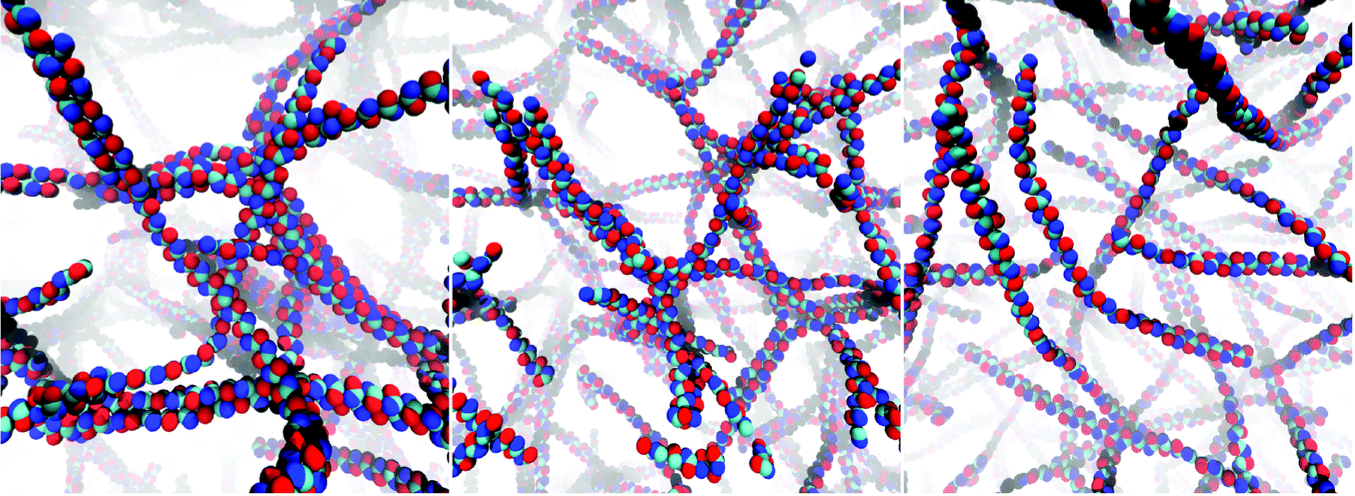}
  \caption{ 
  Typical conformations for a chitosan-based hydrogel
  with a polymer concentration of 8.9\%, 
  and pH of (c) $>10.5$, (d) 6.5, (e) $<2.5$. Red, cyan, and blue beads 
  correspond to chitosan B1, B2, and B3 beads, respectively. 
  Reproduced with permission from Ref.~\citenum{2017HXu-ChemCommun}. 
  Copyright (2017) Royal Society of Chemistry.
  }
  \label{fig:hydrogel}
\end{figure}

For example, \citeauthor{2017HXu-ChemCommun} have developed a Martini model
for chitosan and self-assembled a chitosan hydrogel. Their findings indicate 
that physical crosslinking patterns impact significantly the hydrogel's mechanical properties. 
In particular, increasing the polymer concentration or the pH translates into 
an increase of the elastic modulus of the system, as a consequence of changes 
in the crosslinking patterns (Figure~\ref{fig:hydrogel}).
In another example, using a multi-step protocol, protein imprinting of hydrogels 
was simulated using Martini. Protein imprinting proceeds by polymerizing monomers 
in the presence of a template protein to which monomers are reversibly coordinated. 
Following polymerization the gel is washed leaving - so the idea - specific coordination 
sides for the template protein. Subsequently, by swelling in solution the template 
protein or alike proteins adsorb more preferentially over random 
proteins~\cite{2002MAByrne-AdvDrugDelivRev}. To mimic this process \citeauthor{Zadok2018} 
first simulated coordination of acrylic Martini monomers with lysozyme and cytochrome c. 
Subsequently, using a reaction protocol within LAMMPS, monomers were cross-linked to form a gel. 
After removal of the unreacted monomers a hybrid MD-NVT grand-canonical Monte Carlo simulation 
was used to swell the hydrogel by allowing the water content to change. 
Characterizing the interactions of both proteins with different hydrogels, they found that 
protein binding and selectivity is largely dependent on the nature of the polymer. However, 
it appeared lysozyme overall has the tendency of forming stronger interactions~\cite{Zadok2018}.

\subsection{\normalsize Polymer coatings and glues}
\begin{figure}[htbp]
  \centering
  \includegraphics[width=0.48\textwidth]{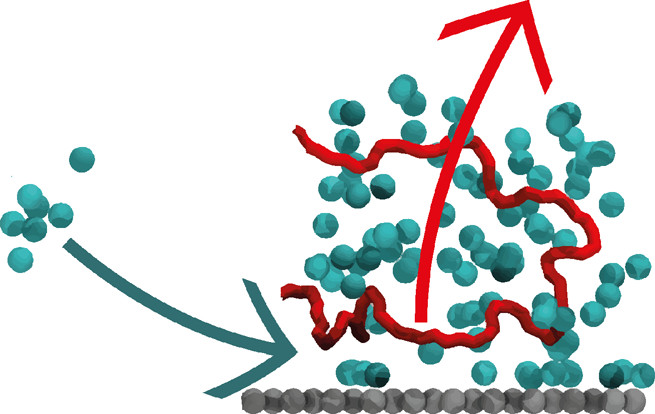}
  \caption{
  Adsorption of solvated polymer chains on a silica surface.~\cite{2018EPerrin-JPhysChemB}
  Polymer solvation was found to play a key role for the adsorption
  of polymers on the silica substrate, highlighting the importance
  of an explicit description of the solvent in such studies.~\cite{2018EPerrin-JPhysChemB}
  Reproduced with permission from Ref.~\citenum{2018EPerrin-JPhysChemB}. 
  Copyright (2018) American Chemical Society.
  }
  \label{fig:polym-on-surfaces}
\end{figure}

Another area where Martini polymers have been used extensively is to study polymer behavior at surfaces and interfaces. Studies have targeted for example polymer conformations at oil/organic solvent-water interfaces~\cite{2014JDong-Langmuir, 2016XQuan-SoftMatter, 2019FJimenez-Angeles-ACSCentSci}, surface water interfaces~\cite{2018EPerrin-JPhysChemB,2011GRossi-Macromolecules}, or even water/air interfaces~\cite{2014SNawaz-JPhysChemB, 2019GCampos-Villalobos-MolSystDesEng}. Polymer behavior at interfaces is interesting in many applications among others for coatings or glues.~\cite{2011GRossi-Macromolecules, 2018EPerrin-JPhysChemB, 2020HVakili-Langmuir} 
For example, \citeauthor{2018EPerrin-JPhysChemB} used Martini to study conformations of PDMA and PAAm absorbed to silica surfaces. The aim of the study was to investigate why PDMA glues to silica whereas PAAm does not regardless of their very similar chemical structure. According to their findings, PAAm is better solvated and therefore does not adhere to silica. They further found that dynamical properties of the polymer close to the surface can only be described by an explicit solvent model 
(Figure~\ref{fig:polym-on-surfaces})~\cite{2018EPerrin-JPhysChemB}.

\subsection{\normalsize Microphase-separated polymers}
\begin{figure}[htbp]
  \centering
  \includegraphics[width=0.48\textwidth]{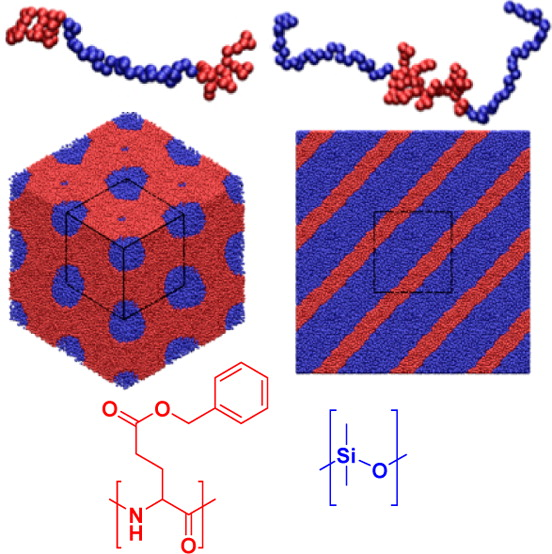}
  \caption{
  Block copolymer morphologies. Hexagonally packed PDMS cylinders
  (left-hand side) and lamellar morphology (right-hand side).
  PBLG is rendered in red, and PDMS in blue.
  Reproduced with permission from Ref.~\citenum{2014JCJohnson-JPhysChemB}. 
  Copyright (2014) American Chemical Society.
  }
  \label{fig:blockcopo}
\end{figure}

Although Martini reproduces well the self-assembly of small surfactants and polymers in aqueous solution~\cite{2012GRossi-JPhysChemB-PEG, 2011MVelinova-Langmuir, 2014JJJanke-Langmuir, Lebard2012, Sangwai2012, 2015SWang-Langmuir, 2017YKChoi-Macromolecules, 2018MVuorte-JPhysChemB,2020HChan-npjComputMater}, self-assembly of large scale polymer - especially block copolymer systems - remains challenging. These micro-phase separated assemblies of copolymers are hugely important in many technological applications. For example, they are a key component for development of next generation batteries~\cite{2014WSYoung-JPolymSciPartBPolymPhys}. 
Hence, simulations of micro-phase separated block copolymers have been
performed with Martini~\cite{2012MGrujicic-JMatSci, 2014JCJohnson-JPhysChemB, 2013MZSlimani-Macromolecules, 2020SLi-PCCP, 2020HVakili-Polymer} 
For example, in a first attempt, \citeauthor{2014JCJohnson-JPhysChemB} managed to self-assemble in an unbiased fashion lamellar, micellar and cylindrical phases of a PDMS-$\gamma$-benzyl-L-glutamate block-copolymer (Figure~\ref{fig:blockcopo}). Their simulations showed a strong correlation between the obtained morphology and the geometry and type of the side chain~\cite{2014JCJohnson-JPhysChemB}. 
In a similar study, \citeauthor{2013MZSlimani-Macromolecules} built a lamellar phase of a polyester gradient co-block-copolymer~\cite{2013MZSlimani-Macromolecules}. These works demonstrate that studying these microphase separated polymer systems is in principle possible with Martini. 

\subsection{\normalsize Block copolymer self-assembly}
A large body of work has has investigated the self-assembly of different block copolymers in water or other 
solvents.~\cite{2015FYuan-JPhysChemB, 
2015FWu-Macromolecules, 
2015WJiang-Langmuir, 
2017DAGrillo-JChemPhys, 
2017X-LSun-JPolSciBPolPhys, 
2018DAGrillo-JChemPhys, 
2018KAKosakowska-Biomacromolecules, 
2018MEElizondo-Garcia-Molecules, 
2019FBrunel-JPhysChemB, 
2020GCampos-Villalobos-JColloidInterfaceSci, 
2020AKoochaki-RSCAdv}
Among these systems, the most studied with Martini are 
poloxamers~\cite{2016UAdhikari-JPhysChemB, 
2016IWood-JMolStruc,   
2018IWood-EurBiophysJ, 
2019JHwang-MacromolRes,
2019JMRAlbano-Processes, 
2019GPerez-Sanchez-JPhysChemC,
2020WWu-Langmuir, 
2020S-MMa-PCCP} 
also known as pluronics, which are PEO-PPO-PEO amphiphilic tri block copolymers.
Such polymers can form a wide range of aggregates ranging
from bilayers, over micelles, to polymersomes, that is, synthetic vesicles,
the latter ones being particularly studied for applications as
nanocarrier devices for drug delivery.
Related to the drug-delivery applications but also of relevance for other technological uses, 
several Martini-based works studied how small molecules
(such as drugs) self-assemble with polymers or diffuse into polymer 
matrices~\cite{2010LXPeng-Biopolymers,2020DPOtto-AAPSPharmSciTech, 
2018AMafi-Biomacromolecules, 2017ELin-ComputMaterSci, 2017ELin-Polymer,
2018GZhang-Macromolecules, 2019HSharma-Macromolecules, 2020DLiu-Polymers}.
In one study, \citeauthor{2019HSharma-Macromolecules} investigating the formation
of polymwer-wrapped and polymer-threaded worm-like micelles as a function
of polymer hydrophobicity and rigidity~\cite{2019HSharma-Macromolecules}.

\begin{figure}[htbp]
  \centering
  \includegraphics[width=0.48\textwidth]{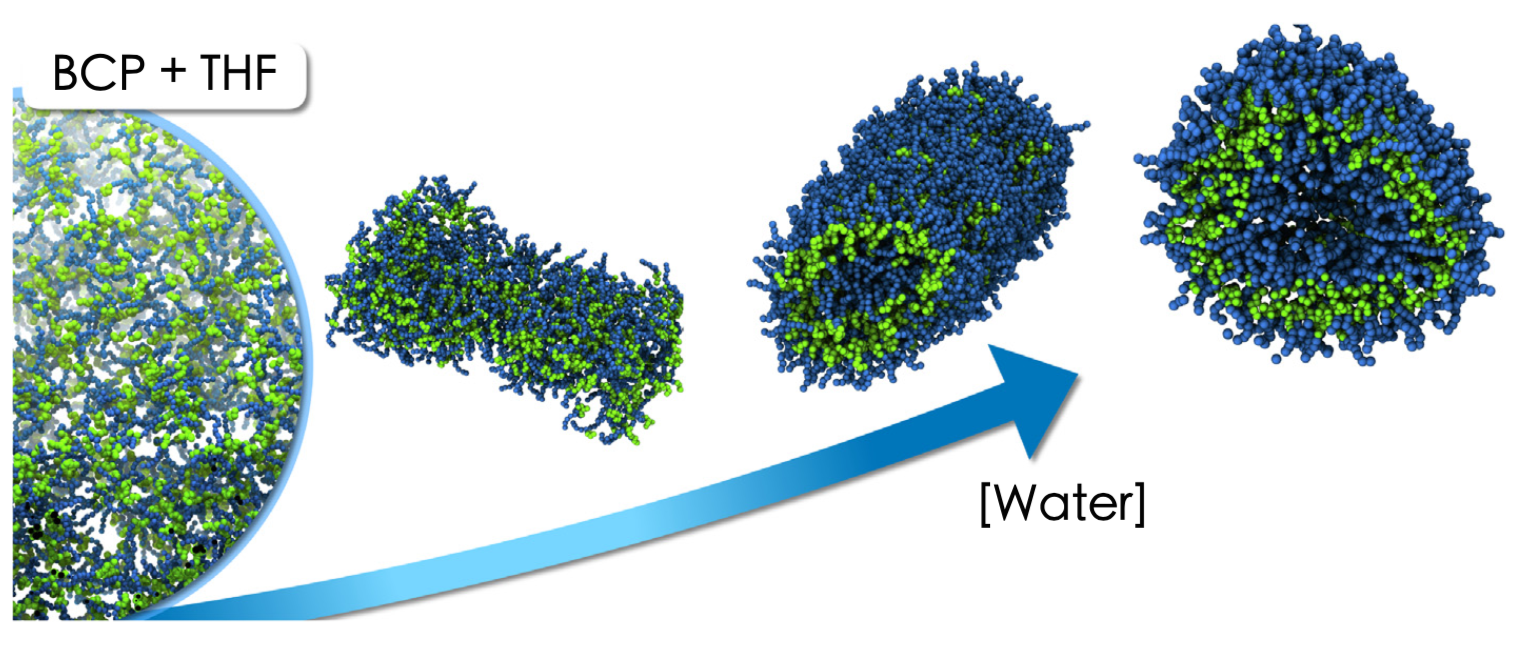}
  \caption{
  Morphology obtained from the self-assembly of PEO-b-PBMA block
  copolymers (BCP) in water and THF
  mixtures~\cite{2020GCampos-Villalobos-JColloidInterfaceSci}.
  The morphology changes from dissolved chains or monomers in THF,
  over dispersed sheets or disk-like aggregates, to vesicles
  as the fraction of water increases. The morphologies are also
  affected by the BCP concentration~\cite{2020GCampos-Villalobos-JColloidInterfaceSci}.
  Reproduced with permission from Ref.~\citenum{2020GCampos-Villalobos-JColloidInterfaceSci}. 
  Copyright (2020) Elsevier.
  }
  \label{fig:blockcopo-in-solution}
\end{figure}

In another study concerning block copolymer self-asssembly, 
\citeauthor{2020GCampos-Villalobos-JColloidInterfaceSci} simulated 
PEO-b-poly(butylmethacrylate) (PBMA) copolymers which are being studied for applications 
as nanostructured materials.~\cite{2020GCampos-Villalobos-JColloidInterfaceSci} 
Martini simulations of the self-assembly of these block copolymers in water and tetrahydrofuran
(THF) mixtures revealed the occurrence of a wide spectrum of mesophases 
(Figure~\ref{fig:blockcopo-in-solution}). The corresponding morphological phase diagram 
of this ternary system includes dispersed sheets or disk-like aggregates, and spherical 
and rod-link vesicles at low block copolymer concentrations, and bicontinuous and lamellar 
phases at high concentrations. Moreover, the THF/water relative content is found to play 
a crucial role on the self-assembly kinetics and resulting 
morphologies~\cite{2020GCampos-Villalobos-JColloidInterfaceSci}.

\subsection{\normalsize Nanoparticles}
Martini parameters are available for a wide variety of nanoparticles 
including fullerenes~\cite{2008JWong-Ekkabut-NatNanotechnol,
2012LMonticelli-JChemTheoryComput, 2009RSGDOrzario-Nanoscale}, 
carbon-nanotubes~\cite{2007EJWallace-NanoLett, 2011NPatra-JAmChemSoc,
2012HLee-JPhysChemC, 2013MLelimousin-Small}, 
and gold~\cite{2011J-QLin-JPhysChemC, 2013JDong-MacromolTheorySimul} 
among others. Apart from their importance to modeling organic electronics,
which are subject to the next section, they have been  used to simulate 
polymer nanoparticle composites~\cite{2012RGUttarwar-IndEngChemRes, 
2013HLee-JPhysChemC-PEG-CNT,  
2013JDong-MacromolTheorySimul,
2014SLin-SoftMatter,          
2014JMaatta-JChemEngData,     
2014JDong-Langmuir,           
2016XQuan-SoftMatter,         
2016OVDeOliveira-ComputTheorChem,
2017HLee-JMolGraphModel,         
2018UDahal-Macromolecules,       
2018SKavyani-JPhysChemB,         
2019PKhan-JPhysChemB,            
2020UDahal-Macromolecules}, 
self-assembly of nanoparticles~\cite{2011J-QLin-JPhysChemC,
2015ZJiang-NatMater, 
2017PFu-ApplThermEng,
2019TKPatra-Nanoscale}, 
and solution processing of nanoparticles~\cite{2020CDWilliams-2DMater, 
2019JLi-AdvMaterInterfaces, 2012DWu-JPhysChemB, 2015C-JShih-JPhysChemC}.

\begin{figure}[htbp]
  \centering
  \includegraphics[width=0.48\textwidth]{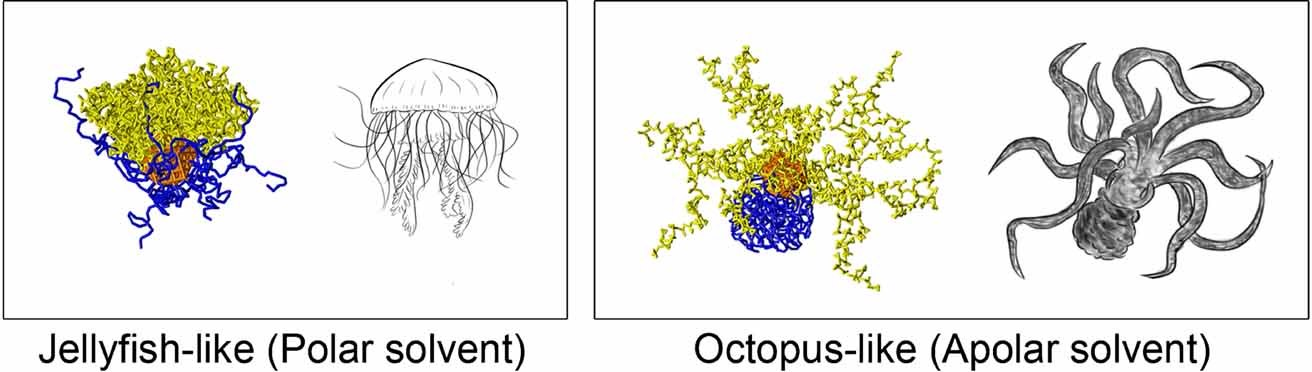}
  \caption{
  Jellyfish-like and octopus-like morphology of gold nanoparticles grafted
  with polymers in a polar and apolar solvent, 
  respectively~\cite{2013JDong-MacromolTheorySimul}.
  Reproduced with permission from Ref.~\citenum{2013JDong-MacromolTheorySimul}. 
  Copyright (2013) Wiley-VCH.
  }
  \label{fig:nanoparticle-polymer}
\end{figure}

For example, the behaviour of PEG grafted covalently or physically onto carbon 
nanotubes has been studied in some detail~\cite{2013HLee-JPhysChemC-PEG-CNT,
2014SLin-SoftMatter, 2014JMaatta-JChemEngData, 2017HLee-JMolGraphModel}. 
In addition to carbon-based nanoparticles, inorganic nanoparticles with 
grafted polymers are an interesting nanomaterial with potential applications in sensoring, 
microfluidics and smart surfaces to name some examples.~\cite{2013JDong-MacromolTheorySimul} 
They are especially interesting for their response to different solvents. 
Within the Martini framework, in particular gold nanoparticles have received  
a lot of attention ~\cite{2013JDong-MacromolTheorySimul, 2014JDong-Langmuir, 
2016XQuan-SoftMatter,2018UDahal-Macromolecules,2020UDahal-Macromolecules}. 
For example, \citeauthor{2013JDong-MacromolTheorySimul} have studied 
the solvent behavior of differently composed PEO-b-PS block copolymers attached 
to gold nanoparticles. They found varying the composition of the block copolymer 
leads to a variety of different morphologies. Some of them were the expected 
sphere-shell like conformations where the polymers extend or collapse in a trivial 
fashion onto the nanoparticles. On the other hand, \citeauthor{2013JDong-MacromolTheorySimul} 
also identified some non-trivial conformations described as rings, buckles, 
and sectorially arranged chains 
(Figure~\ref{fig:nanoparticle-polymer})~\cite{2013JDong-MacromolTheorySimul}. 
\citeauthor{2018UDahal-Macromolecules} have studied in detail PEO-grafted gold 
nanoparticles, investigating hydration and structural properties as a function 
of PEO chain length and grafting density~\cite{2018UDahal-Macromolecules, 
2020UDahal-Macromolecules}, with observed properties in agreement with experimental 
data but providing a microscopic view on such nanoparticle-polymer composites. 
These examples highlight the possibility of using Martini to scan many different 
compositions for such systems and optimize a target behaviour without 
the need for experimentally synthesizing  all of the structures.  
The study of materials composed of self-assembled coated nanoparticles
is another interesting effort. Solids formed by such nanoparticles form 
a versatile class of hybrid materials in which both the nanoparticle core 
and the organic ligand shell can be tuned, leading to a variety of materials. 
For example, Martini as been used to simulate single-layer coated nanoparticle 
membranes~\cite{2015ZJiang-NatMater},
or nanoparticle superlattices for energy applications~\cite{2019TKPatra-Nanoscale}.
Additionally, the dispersion of CNTs and 2D materials, such as graphene and MXene, 
in surfactant aqueous solutions~\cite{2012DWu-JPhysChemB,2015C-JShih-JPhysChemC, 
2020CDWilliams-2DMater,2019JLi-AdvMaterInterfaces} has been investigated.
understanding and optimizing such dispersion is of paramount
importance for the processing of these materials; in the case of the
2D material MXene, \citeauthor{2019JLi-AdvMaterInterfaces} probed
three typical surfactants with different structural characteristics,
finding that the surfactant with long hydrocarbon chain and
positively charged head group can form stable bilayers
at the surface with MXene, which has implications fot the thermal energy
dissipation of the 2D material~\cite{2019JLi-AdvMaterInterfaces}.
Other applications in the field of nanoparticle composite materials include 
polymer/graphene~\cite{2019MShi-JApplPhys}, polymer/graphite~\cite{2019MShi-JApplPolymSci}, 
polymer/clay~\cite{2019PKhan-JPhysChemB} composites,
and polymer-CNT-protein matrices for applications in the field of tissue
regeneration~\cite{2019MMSlepchenkov-Materials}.

\subsection{\normalsize Organic electronics}
The morphology of the organic material which constitutes the 
active layer of organic electronic devices is
a critical parameter for the functioning of such devices.
Computational modeling of the morphology represents a fundamental 
step towards an increased rational approach to the design of 
high-performance organic materials for electronic 
applications.~\cite{2018GHan-AdvEnergyMater,2019PFriederich-AdvMater}
It is possible to model the morphology of organic electronic materials with Martini, 
in particular to obtain and characterize morphologies, which are often composed of
more than one organic semiconductor~\cite{2016TWinands-PhysChemChemPhys,
2017RAlessandri-JAmChemSoc,2017LQiu-JMaterChemA, 2018JMunshi-ComputMaterSci, 
2019JMunshi-ACSApplMaterInterfaces, 2019JMunshi-JPolSciBPolPhys,
2020JMunshi-SoftMatter, 2019JLee-JMaterChemA};
and to subsequently backmap~\cite{2014TAWassenaar-JChemTheoryComput} 
the obtained CG
morphologies to atomistic resolution, a step often useful in order to perform 
fine-grained calculations aimed at evaluating the electronic properties of such
materials~\cite{2015MBockmann-PhysChemChemPhys, 2016TWinands-PhysChemChemPhys,
2017RAlessandri-JAmChemSoc, 2020RAlessandri-AdvFunctMater,
2020NRolland-ComputMaterSci}.
Martini models have been already developed for many prototypical organic 
semiconductors used in organic electronic devices, such as conjugated 
polymers~\cite{2017RAlessandri-JAmChemSoc,
2018MModarresi-PhysChemChemPhys, 2017YKChoi-Macromolecules}, small conjugated 
molecules~\cite{2016TWinands-PhysChemChemPhys, 2017LQiu-JMaterChemA, 
2018JLiu-AdvMater, 2020JGJang-ACSApplMaterInterfaces}, and C$_{60}$
fullerene~\cite{2008JWong-Ekkabut-NatNanotechnol, 2012LMonticelli-JChemTheoryComput} 
and some of its derivatives~\cite{2017RAlessandri-JAmChemSoc, 2017LQiu-JMaterChemA,
2020RAlessandri-AdvFunctMater}.
Arguably one of the most popular subfields of organic electronics 
is organic photovoltaics. Systems such as 
P3HT:DiPBI~\cite{2016TWinands-PhysChemChemPhys},
P3HT:PCBM~\cite{2017RAlessandri-JAmChemSoc, 2018JMunshi-ComputMaterSci, 
2019JMunshi-ACSApplMaterInterfaces, 2019JMunshi-JPolSciBPolPhys,
2020JMunshi-SoftMatter, 2021JMunshi-ComputMaterSci}, PBDB-T:F-ITIC~\cite{2019JLee-JMaterChemA}, 
and P3HT:PTEG-1~\cite{2020RAlessandri-AdvFunctMater}
have already been simulated with Martini.
Simulations of neat P3HT~\cite{2015MBockmann-PhysChemChemPhys} have also
been performed, while more organic semiconductor mixtures have been tested
in the context of organic thermoelectric devices~\cite{2017LQiu-JMaterChemA, 
2018JLiu-AdvMater} and organic mixed ion-electron conductors (which will be
described as part of the next section)~\cite{2018MModarresi-PhysChemChemPhys, 
2019MModarresi-PhysChemChemPhys,2019JRehmen-ACSOmega, 
2020NRolland-ComputMaterSci,2020MModarresi-Macromolecules}.

\begin{figure}[htbp]
  \centering
  \includegraphics[width=0.48\textwidth]{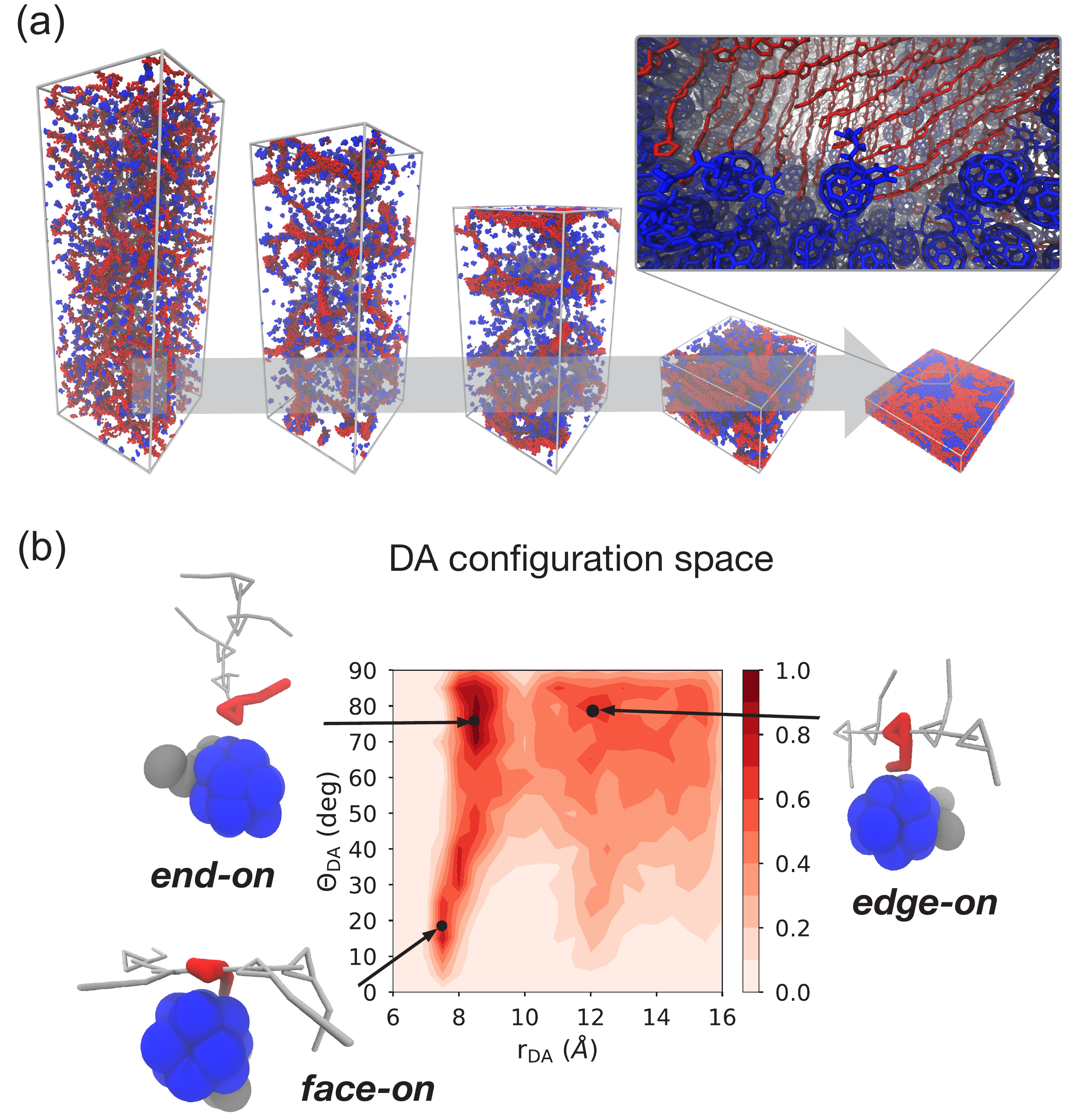}
  \caption{
  Bulk heterojunction morphologies from solvent evaporation simulations 
  for a P3HT-PCBM mixture~\cite{2017RAlessandri-JAmChemSoc}.
  (a) The inset shows the resulting atomistic structure obtained via backmapping.
  P3HT polymer chains are rendered in red and PCBM molecules in blue, respectively.
  (b) Molecular orientations at the donor-acceptor (DA) interface can be resolved,
  also as a function of molecular features and processing 
  conditions~\cite{2020RAlessandri-AdvFunctMater}.
  (a) Reproduced with permission from Ref.~\citenum{2017RAlessandri-JAmChemSoc}. 
  Copyright (2017) American Chemical Society.
  (b) Reproduced with permission from Ref.~\citenum{2020RAlessandri-AdvFunctMater}. 
  Copyright (2020) Wiley-VCH.
  }
  \label{fig:P3HTPCBM-solvent-evap}
\end{figure}

When modeling the morphology of organic electronic materials,
simulating fabrication processes, such as solution-processing and thermal annealing,
is an important step to be taken into account and which can be studied to obtain in silico insights.
The solvent evaporation process which takes place during the fabrication of
organic thin films can be simulated 
by simulating ``bulk'' evaporation, as first shown by 
\citeauthor{2014CKLee-JPhysChemC} using a supra CG model.~\cite{2014CKLee-JPhysChemC}
Similar solvent evaporation simulations have been applied to 
simulate the prototypical polymer:fullerene photovoltaic
blend--P3HT:PCBM--at the Martini level 
by \citeauthor{2017RAlessandri-JAmChemSoc}
(Figure~\ref{fig:P3HTPCBM-solvent-evap})~\cite{2017RAlessandri-JAmChemSoc}.
Other Martini organic semiconductor thin films have been
solution-processed in silico~\cite{2018JMunshi-ComputMaterSci, 
2018MModarresi-PhysChemChemPhys, 2019JMunshi-ACSApplMaterInterfaces, 
2019JMunshi-JPolSciBPolPhys, 2020NRolland-ComputMaterSci,
2020RAlessandri-AdvFunctMater, 2020JMunshi-SoftMatter, 2021JMunshi-ComputMaterSci}.
In particular, \citeauthor{2017RAlessandri-JAmChemSoc} studied the evolution of the
morphology of P3HT:PCBM blends as a function of the molecular weight of P3HT, 
the solvent evaporation rate, and thermal annealing.
In agreement with experiments, thermal annealing and slower evaporation rates
lead to larger phase separation and increased crystallinity of the P3HT
phase. The crystallinity of P3HT could be probed by computing scattering
signals, which were found to be in qualitative agreement with experimental
data\cite{2017RAlessandri-JAmChemSoc}.
The too-large size of S-beads, however, prevents a quantitative reproduction of the 
stacking distance between the polythiophene backbones~\cite{2017RAlessandri-JAmChemSoc}. 
Besides allowing to quantify the degree of crystallinity, computing
X-ray scattering signals of simulated morphologies 
allows for comparison to experimental data.
Other works made comparison between CG and X-ray scattering
data~\cite{2017RAlessandri-JAmChemSoc, 2019JMunshi-ACSApplMaterInterfaces,
2019AYMehandzhiyski-ACSApplEnergyMater, 2020NRolland-ComputMaterSci},
or to scanning electron microscopy~\cite{2017RAlessandri-JAmChemSoc}
or atomic force microscopy (AFM)~\cite{2017LQiu-JMaterChemA,
2019AYMehandzhiyski-ACSApplEnergyMater} images.
Martini simulations allow to scan a range of parameters which are known
to affect the morphology of organic thin films in the lab,
such as weight ratio of the components\cite{2019JMunshi-ACSApplMaterInterfaces}, 
polymer polydispersity~\cite{2019JMunshi-JPolSciBPolPhys}, 
molecular weight~\cite{2017RAlessandri-JAmChemSoc,2019JMunshi-ACSApplMaterInterfaces},
post-~\cite{2017RAlessandri-JAmChemSoc,2019JMunshi-ACSApplMaterInterfaces} and
pre-~\cite{2019JMunshi-ACSApplMaterInterfaces}-evaporation heating treatments.
Moreover, once morphologies have been generated, their macroscopic properties, such
as mechanical properties~\cite{2020JMunshi-SoftMatter}, or microscopic features,
such as the molecular orientations at the interfaces between the two blended
organic semiconductors~\cite{2020RAlessandri-AdvFunctMater} 
(Figure~\ref{fig:P3HTPCBM-solvent-evap}b), can be investigated, possibly
as a function of the above parameters.
Finally, other works have looked at the solubility of small molecules used
as dopants in environments of different polarity\cite{2018JLiu-AdvMater}, 
another application which is suited to the Martini model.

An important advantage of obtaining morphologies at the Martini level
is the possibility of directly backmapping~\cite{2014TAWassenaar-JChemTheoryComput} 
the CG morphologies to atomistic
resolution (Figure~\ref{fig:P3HTPCBM-solvent-evap}, inset), 
hence obtaining atom-resolved structures which take into account the self-organization
process which occurs during the processing of an organic 
blend~\cite{2017RAlessandri-JAmChemSoc}.
Indeed, 
large-scale morphologies have been backmapped
to atomistic resolution in order to compute, by means of (semi-empirical) quantum
chemical calculations: 
UV/Vis spectra~\cite{2015MBockmann-PhysChemChemPhys,2016TWinands-PhysChemChemPhys},
energy levels taking into account the local molecular
environment~\cite{2020RAlessandri-AdvFunctMater},
and charge carrier hopping rates~\cite{2020NRolland-ComputMaterSci} 
for charge transport calculations.

\subsection{\normalsize Ion-conducting organic materials}
\begin{figure}[htbp]
  \centering
  \includegraphics[width=0.48\textwidth]{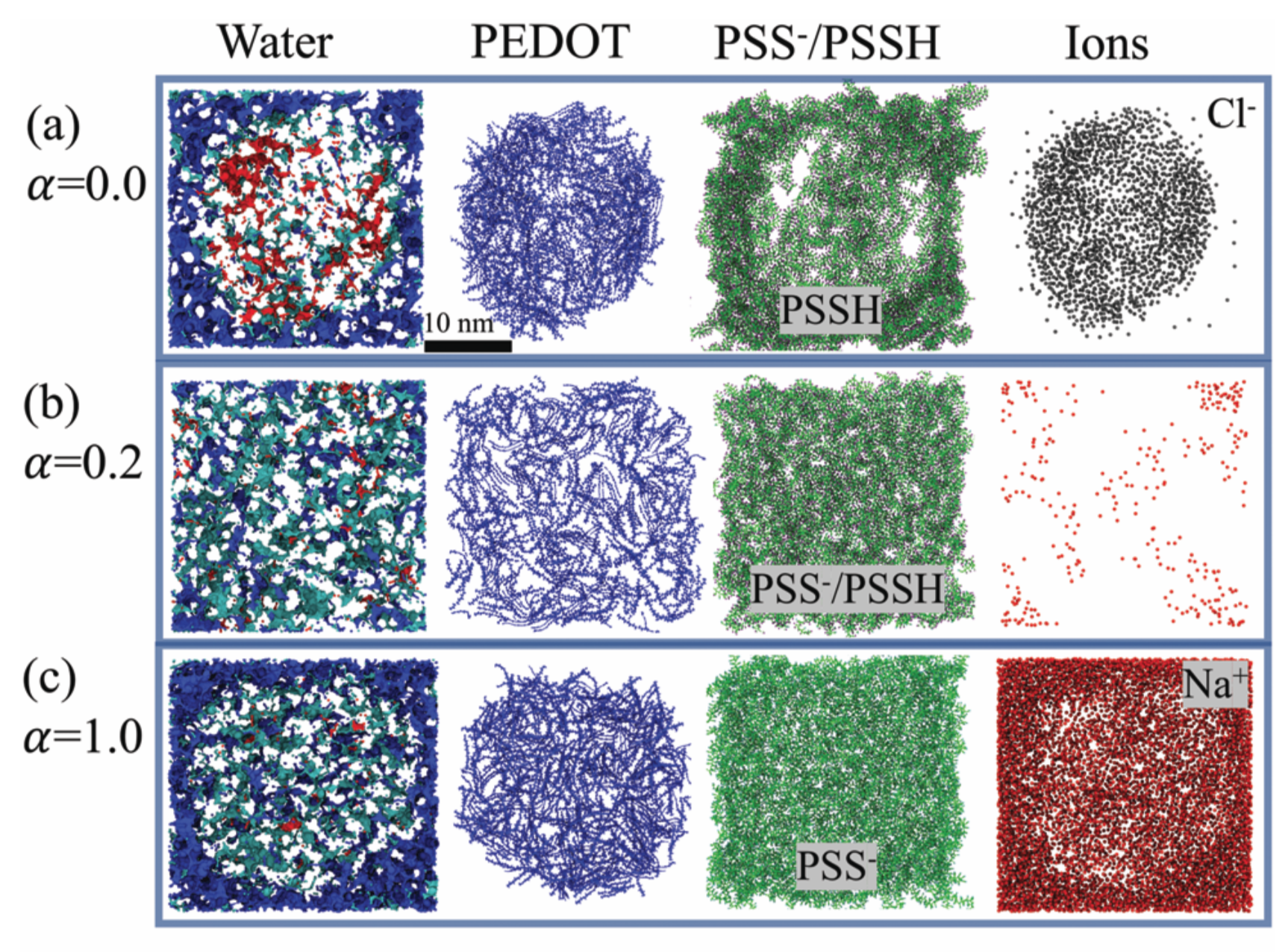}
  \caption{
  Representative morphologies of PEDOT:PSS systems as a function
  of the (PSS) deprotonation level, $\alpha$, which is a function of
  the pH.~\cite{2019MModarresi-PhysChemChemPhys}
  For $\alpha=0$ all PSS chains are protonated (PSSH), 
  vice versa for $\alpha=1$.
  The water phase (left-hand side) is colored in red, blue, or cyan,
  when the water molecules are within a distance of 6 \AA $\:$ from PEDOT,
  PSS, or both, respectively.
  Reproduced with permission from Ref.~\citenum{2019MModarresi-PhysChemChemPhys}. 
  Copyright (2019) Royal Society of Chemistry.
  }
  \label{fig:PEDOT-PSS-morphologies}
\end{figure}

Martini has also been used to investigate~\cite{2018MModarresi-PhysChemChemPhys, 
2019MModarresi-PhysChemChemPhys, 2019JRehmen-ACSOmega, 2020NRolland-ComputMaterSci,
2020MModarresi-Macromolecules, 2021KJain-JColloidInterfaceSci}
organic mixed ion-electron conductors
which are soft (semi-)conductors---often polymers---that
readily solvate and transport ionic 
species~\cite{2019MBerggren-AdvMater, 2019BDPaulsen-NatMater}. 
Applications of such systems include organic electrochemical transistors 
for biological interfacing and neuromorphic devices, 
among others~\cite{2018JRivnay-NatRevMater}.
A Martini model exists for the workhorse system of this field: 
poly(3,4-ethylenedioxythiophene):polystyrene sulfonate 
(PEDOT:PSS)~\cite{2019MModarresi-PhysChemChemPhys}.
Modarresi, Zozoulenko, and co-workers spearheaded Martini-based studies in this 
field by developing a model for
PEDOT and investigating ion diffusion in morphologies of PEDOT:Tos, 
a system made of PEDOT chains and tosylate (Tos), a negatively charged 
molecular  counterion.~\cite{2018MModarresi-PhysChemChemPhys}
The same authors, combining PEDOT with the available~\cite{2015MVogele-JChemPhys}
PSS model, went on to investigate in detail PEDOT:PSS 
morphologies~\cite{2019MModarresi-PhysChemChemPhys, 2020MModarresi-Macromolecules}. 
The simulations allowed to study the effect of pH on the morphology
of in silico solution-processed PEDOT:PSS thin films.
Changes in pH were found to greatly affect the morphology (Figure~\ref{fig:PEDOT-PSS-morphologies}),
and in turn the distribution of the 5-15 weight \% of water content
in the polymer film, which is of critical importance for ion diffusion.
%
Once again, the possibility offered by Martini of easily combining models
allowed \citeauthor{2019AYMehandzhiyski-ACSApplEnergyMater} to
simulate PEDOT:PSS/cellulose~\cite{2011JWohlert-JChemTheoryComput} 
composite paper~\cite{2019AYMehandzhiyski-ACSApplEnergyMater}; such
paper can be used in applications such as 
fuel cells, sensors, and batteries, among others~\cite{2016AMalti-AdvSci}.
The authors could pinpoint the most likely configuration of PEDOT and PSS/PSSH
chains around cellulose by comparing the simulation results to AFM
images~\cite{2019AYMehandzhiyski-ACSApplEnergyMater}. They could identify
the most likely morphology observed in the experiments, namely a bead-like structure
caused by PEDOT aggregates on the fibril which are separated by regions with a lower 
density.~\cite{2019AYMehandzhiyski-ACSApplEnergyMater}

The works presented above on mixed ion-electron conductors partly build
on earlier developments of Martini models for polyelectrolytes,
which include models for polystyrenesulfonate (PSS)~\cite{2015MVogele-JChemPhys, 
2015SMantha-JPhysChemB}, poly(diallyldimethylammonium)
(PDADMA)~\cite{2015MVogele-JChemPhys} and, more recently,
partially hydrolyzed polyacrylamide (HPAM)~\cite{2018SHu-ComputMaterSci}.
In the case of such highly charged systems, polarizability introduced for water and other
beads has been shown to be important~\cite{2010SOYesylevskyy-PLoSComputBiol,
2017JMichalowsky-JChemPhys, 2015MVogele-JMolLiq, 2018JMichalowsky-JChemPhys,
2013DHdeJong-JChemTheoryComput}. Hence, when charged interactions are expected
to be important, it is recommended to develop models in the context of the
polarizable~\cite{2010SOYesylevskyy-PLoSComputBiol, 2017JMichalowsky-JChemPhys} 
water model.
For example, \citeauthor{2015MVogele-JChemPhys} have shown that the Martini model of PSS 
used in polarizable water is able to accurately describe ion distributions 
around the polymer and reduction in dielectric screening. These important 
properties are not reproduced in regular Martini water.~\cite{2015MVogele-JChemPhys}

\begin{figure}[htbp]
  \centering
  \includegraphics[width=0.48\textwidth]{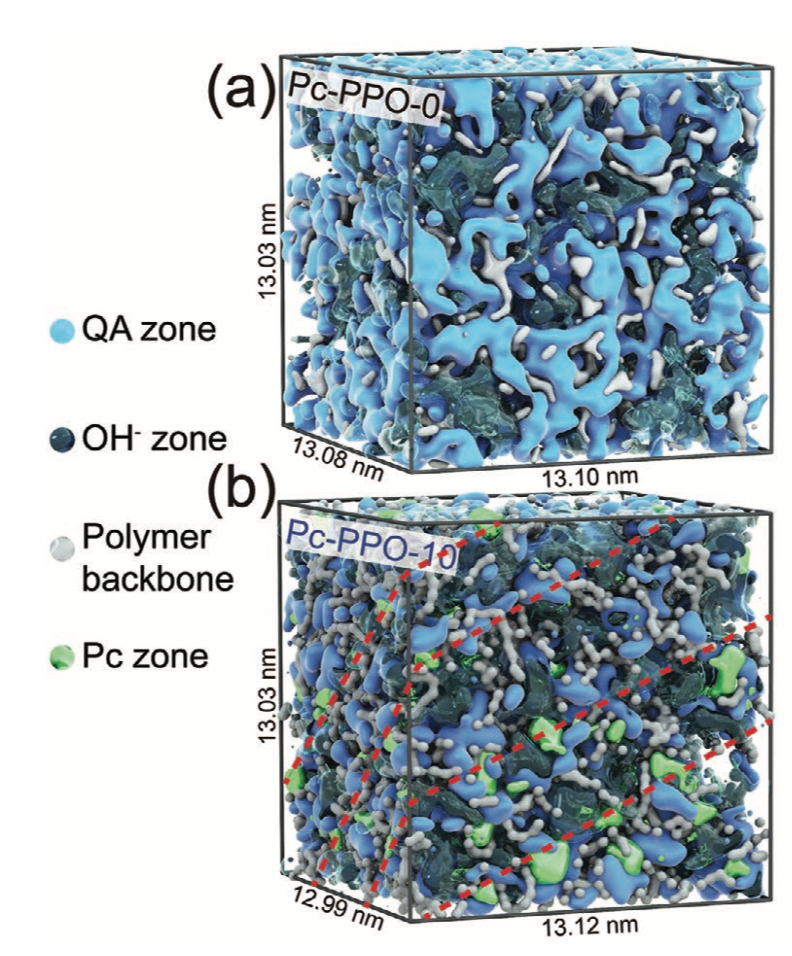}
  \caption{
  Martini models of anion exchange membranes for fuel 
  cells~\cite{2020YYang-AdvMaterInterfaces}.
  Introduction of (a) 10\% of quaternary ammonium (QA) phthalocyanine (Pc) groups
  (Pc-PPO-10) induces a more structured self-assembly than the 
  (b) random morphology formed by Pc-PPO-0 (where QA-Pc groups are not present), 
  leading to enhanced hydroxide (OH$^{-}$) conductivity.
  Here, PPO stands for poly(2,6-dimethyl-1,4-phenylene oxide).
  Reproduced with permission from Ref.~\citenum{2020YYang-AdvMaterInterfaces}. 
  Copyright (2020) Wiley-VCH.
  }
  \label{fig:anion-exchange-membranes}
\end{figure}

Furthermore,
Martini has been used to model ion-conducting polymeric materials used
as ion exchange membranes for fuel cell applications~\cite{2008KMalek-JChemPhys,
2019WGoncalves-JPhysChemC, 2014JPan-EnergyEnvironSci, 2020TMabuchi-JPolymSci, 
2020SChen-JMembrSci, 2020YYang-AdvMaterInterfaces, 2020TMabuchi-Macromolecules}.
Proton exchange membrane fuel cells use acid 
polyelectrolytes, such as Nafion, as the membrane material. 
The membranes are formed by solution processing techniques. 
In this context,
\citeauthor{2020TMabuchi-JPolymSci} investigated dilute solutions 
of Nafion ionomers, the initial stage of solution processing:
they studied ionomer self-assembly in mixtures of 1-propanol and water
and probed the effects of ionomer concentration, alcohol content, and 
inclusion of salt~\cite{2020TMabuchi-JPolymSci, 2020TMabuchi-Macromolecules}.
\citeauthor{2019WGoncalves-JPhysChemC} studied instead how cavities nucleate 
and grow in a hydrated Nafion membrane subject to mechanical 
deformation, obtaining a nanoscale view on the mechanical properties
of such membranes~\cite{2019WGoncalves-JPhysChemC}.
Next to proton exchange membranes, also
alkaline anion-exchange membranes, which transport instead alkaline 
anions (usually hydroxide), have been modeled with Martini~\cite{2014JPan-EnergyEnvironSci,
2020YYang-AdvMaterInterfaces,2020SChen-JMembrSci}.
\citeauthor{2014JPan-EnergyEnvironSci}, for instance, screened for 
different structural designs of 
polymer electrolytes which 
would increase hydroxide mobility, a key performance parameter.
The prediction was implemented experimentally and found to lead 
to increased hydroxide mobility, which reached efficiencies as high as the 
proton mobility in the more developed Nafion-based proton exchange 
membranes~\cite{2014JPan-EnergyEnvironSci}. 
Finally, recently 
\citeauthor{2020YYang-AdvMaterInterfaces} used the Martini model to microscopically
investigate the impact of adding quaternary ammonium phthalocyanine (Pc) groups
into anion exchange membranes based on poly(2,6-dimethyl-1,4-phenylene oxide) 
for alkaline fuel cells~\cite{2020YYang-AdvMaterInterfaces}.
The self-assembly of Pc helps structuring the anion exchange membrane, 
increasing its hydroxide conductivity 
(Figure~\ref{fig:anion-exchange-membranes})~\cite{2020YYang-AdvMaterInterfaces}.

\subsection{\normalsize Self-assembled supramolecular materials}
Self-assembling of molecular building blocks into supramolecular 
materials holds much promise for a range of potential applications 
in nanotechnology.~\cite{2012TAida-Science,2018PWJMFrederix-ChemSocRev} 
Molecular building blocks which are very popular are short peptides 
(2--10 residues) and peptide conjugates, which can give rise to a large 
variety of biocompatible nanostructures~\cite{2018ALampel-ChemSocRev}.
Many groups have explored their self-assembly process by leveraging the 
Martini model~\cite{2011PWJMFrederix-JPhysChemLett,2012CGuo-ACSNano,
2012MSeo-JChemTheoryComput,2012O-SLee-NanoLett, 2013MMazza-ACSNano,
2013DRLewis-Biomaterials,
2013TYu-JPhysChemB,2014NThota-RSCAdv, 2015MRad-Malekshahi-JAmChemSoc, 
2015PWJMFrederix-NatChem, 2015MSlyngborg-PCCP, 2016GGScott-AdvMater,
2016JKwon-JComputChem, 2017FSun-JPhysChemB,2017IRSasselli-OrgBiomolChem,
2018NLIng-ACSNano,
2018RPugliese-ActaBiomater,2018NBrown-ACSNano,2018SMushnoori-OrgBiomolChem,
2020YWang-MaterChemFront, 2020CYJLau-CommunChem, 2020ERDraper-Matter}. 
Besides allowing to simulate self-assembly and growth, 
Martini is also particularly suited for high-throughput applications.
An example of such a high-throughput application in the area of peptide-based 
supramolecular materials is the work of \citeauthor{2015PWJMFrederix-NatChem} 
who simulated all 8000 combinations of tripeptides~\cite{2015PWJMFrederix-NatChem}. 
The prediction of self-assembling and non-assembling peptide sequences coming from 
the Martini simulations were verified by a full experimental characterization, showing
the predictive power of Martini in this area. The approach allowed to extract 
guidelines for new peptide materials~\cite{2015PWJMFrederix-NatChem}. 
The modularity of the Martini model also allows to easily combine peptides with other 
molecular moieties. Accordingly, \citeauthor{2017RAMansbach-JPhysChemB} built 
models for $\pi$-conjugated peptides, which are promising bioelectronic materials 
due to their optoelectronic properties, and studied their self-assembly in 
detail~\cite{2017RAMansbach-JPhysChemB,
2017RAMansbach-OrgBiomolChem, 2018RAMansbach-JPhysChemB, 2020KShmilovich-JPhysChemB}.

\begin{figure}[htbp]
  \centering
  \includegraphics[width=0.48\textwidth]{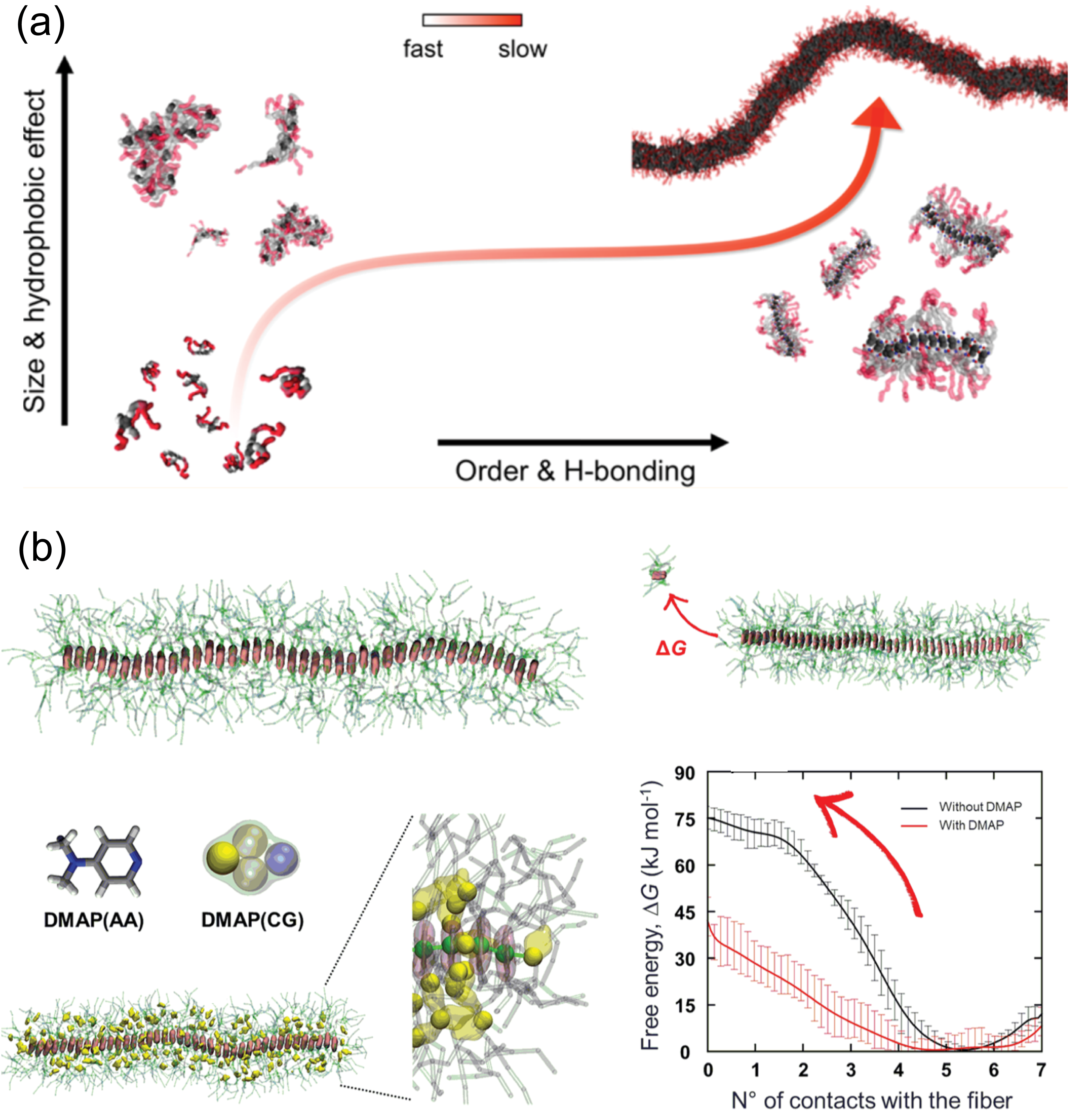}
  \caption{
  Examples of supramolecular polymers.
  (a) Formation of BTA supramolecular polymers, as resolved
  by Martini CG simulations,
  proceeds via an initial fast aggregation followed by a slower
  reorganization and fiber growth.~\cite{2017DBochicchio-ACSNano}.
  (b) Modeling of porphyrin-based supramolecular copolymers:
  the presence of DMAP small-molecules eases porphyrin monomer exchange 
  in and out the fiber, as quantified by the free energy 
  profiles~\cite{2018SHJung-JAmChemSoc}.
  (a) Reproduced with permission from Ref.~\citenum{2017DBochicchio-ACSNano}. 
  Copyright (2017) American Chemical Society.
  (b) Reproduced with permission from Ref.~\citenum{2018SHJung-JAmChemSoc}. 
  Copyright (2017) American Chemical Society.
  }
  \label{fig:supramolecular-BTA-and-porphyrin}
\end{figure}

Besides peptide-based compounds, studies on other supramolecular systems have
been reported in more recent years.
An important example are 
1,3,5-benzenetricarboxamide (BTA)-based supramolecular polymers,
synthetic supramolecular materials
which have been studied in detail by Bochicchio, Pavan, and 
co-workers~\cite{2017DBochicchio-ACSNano, 2017DBochicchio-NatCommun,
2017DBochicchio-JPCL-DRY-BTA, 2018ATorchi-JPhysChemB,
2019RPMLafleur-Macromolecules, 2020PGasparotto-JPhysChemB, 2020SVarela-Aramburu-Biomacromolecules}. 
Two Martini models for the BTA core were developed, both capturing the step-wise 
cooperative polymerization mechanism which leads to the formation of supramolecular 
fibers (Figure~\ref{fig:supramolecular-BTA-and-porphyrin})~\cite{2017DBochicchio-ACSNano}: 
BTA monomers initially aggregate quickly in water due to hydrophobic interactions;
the disordered aggregates formed then reorganize into directional oligomers on a slower
time scale; such ordered oligomers then fuse to form and elongate the supramolecular
fiber on an even slower time scale. 
In the more refined model, additional charged particles were introduced to 
improve the stacking interactions between the BTA 
cores~\cite{2017DBochicchio-ACSNano}, similarly to the ones introduced
within amino acid side chains by \citeauthor{2013DHdeJong-JChemTheoryComput} in the 
polarizable version of the Martini protein model.~\cite{2013DHdeJong-JChemTheoryComput}
The refined model allows for accurate monitoring of hydrogen-bonding between
the BTA monomers, while the simpler model---where such charges are omitted---is 
recommended for studies of interactions of BTA-based assemblies with other 
(macro)molecules~\cite{2017DBochicchio-ACSNano}.
A Dry Martini\cite{2015CArnarez-JChemTheoryComput} version of the BTA model 
has also been put forward by the same authors~\cite{2017DBochicchio-JPCL-DRY-BTA}. 
There are several applications of structural variants of the BTA supramolecular
polymer~\cite{2017DBochicchio-NatCommun, 2018ATorchi-JPhysChemB,
2019RPMLafleur-Macromolecules, 2020PGasparotto-JPhysChemB},
some in combination with enhanced sampling and machine-learning
techniques~\cite{2017DBochicchio-NatCommun, 2020PGasparotto-JPhysChemB}.
For example,
Martini-based well-tempered metadynamics simulations were used to
investigate monomer exchange in and out of the fibers, finding a central role of
defects on the supramolecular structure in this process~\cite{2017DBochicchio-NatCommun}.
Being able to characterize these defects may thus be important to control the
dynamic behavior and properties of such systems: \citeauthor{2020PGasparotto-JPhysChemB}
used machine-learning techniques to systematically identify and compare such defects 
in this class of supramolecular materials\cite{2020PGasparotto-JPhysChemB}.

Other supramolecular polymer aggregates studied include:
benzotrithiophene (BTT)-based supramolecular fibers~\cite{2018NMCasellas-ChemCommun},
azobenzene-containing monomers which assemble in a supramolecular
tubule~\cite{2019DBochicchio-ACSNano}, and 
porphyrin-based supramolecular polymers~\cite{2018SHJung-JAmChemSoc}.
In the latter study, Martini-based well-tempered metadynamics 
allowed \citeauthor{2018SHJung-JAmChemSoc} to quantify the 
effect of a small molecule, DMAP, on the monomer exchange from the fiber 
(Figure~\ref{fig:supramolecular-BTA-and-porphyrin}b)~\cite{2018SHJung-JAmChemSoc}.
The simulations showed that DMAP molecules interfere
with the monomer-monomer interactions at the fiber ends by first penetrating in
between the monomer porphyrin cores and then 
facilitating monomer dissociation from the fiber end~\cite{2018SHJung-JAmChemSoc}.
Other supramolecular material systems which have been simulated
with Martini include supramolecular block copolymers~\cite{2020ASarkar-JAmChemSoc-A,
2020ASarkar-JAmChemSoc-B}, 
peptide-based supramolecular polymers chemically linked to
spiropyran-based networks~\cite{2020CLi-NatMater},
poly-catenanes~\cite{2020SDatta-Nature}, 
peptoid-based nanomaterials~\cite{2020MZhao-JPhysChemB}, 
supramolecular macrocycle fibers~\cite{2020SMaity-JAmChemSoc},
responsive conjugated polymers~\cite{2019YKChoi-JMaterChemC},
and light-harvesting double-walled nanotubes~\cite{SOONIPatmanidis-CGing-C8S3-submitted}.

\subsection{\normalsize Green solvents}
Historically, the Martini model has been developed for the simulation of biological membranes formed by lipids. With lipids being one special class of surfactants early on Martini was extended to simulate the assembly and interaction of other synthetic ionic and non-ionic surfactants.~\cite{2012GRossi-JPhysChemB-PEG, 2016APizzirusso-JPhysChemB, 2015SWang-Langmuir, 2012DSergi-JChemPhys, 2019SDAnogiannakis-JPhysChemB, 2020SDPeroukidis-JChemTheoryComput, 2020HBastos-PCCP, 2020GPerez-Sanchez-JPhysChemB, 2020TDPotter-SoftMatter} Recently, interest in surfactants has renewed as ionic liquids (ILs) have attracted much attention for their use as bio-compatible and green solvents and co-solvents. This has led several authors to use Martini to simulate the self-assembly of IL mesophases~\cite{2020EACrespo-JColloidInterfaceSci, 2020LIVazquez-Salazar-GreenChem}, the process of IL mediated extractions~\cite{2020LIVazquez-Salazar-GreenChem, 2020GHuet-GreenChem} as well as to guide the design of de novo molecules~\cite{2020GHuet-GreenChem}.
\begin{figure}[htbp]
  \centering
  \includegraphics[width=0.46\textwidth]{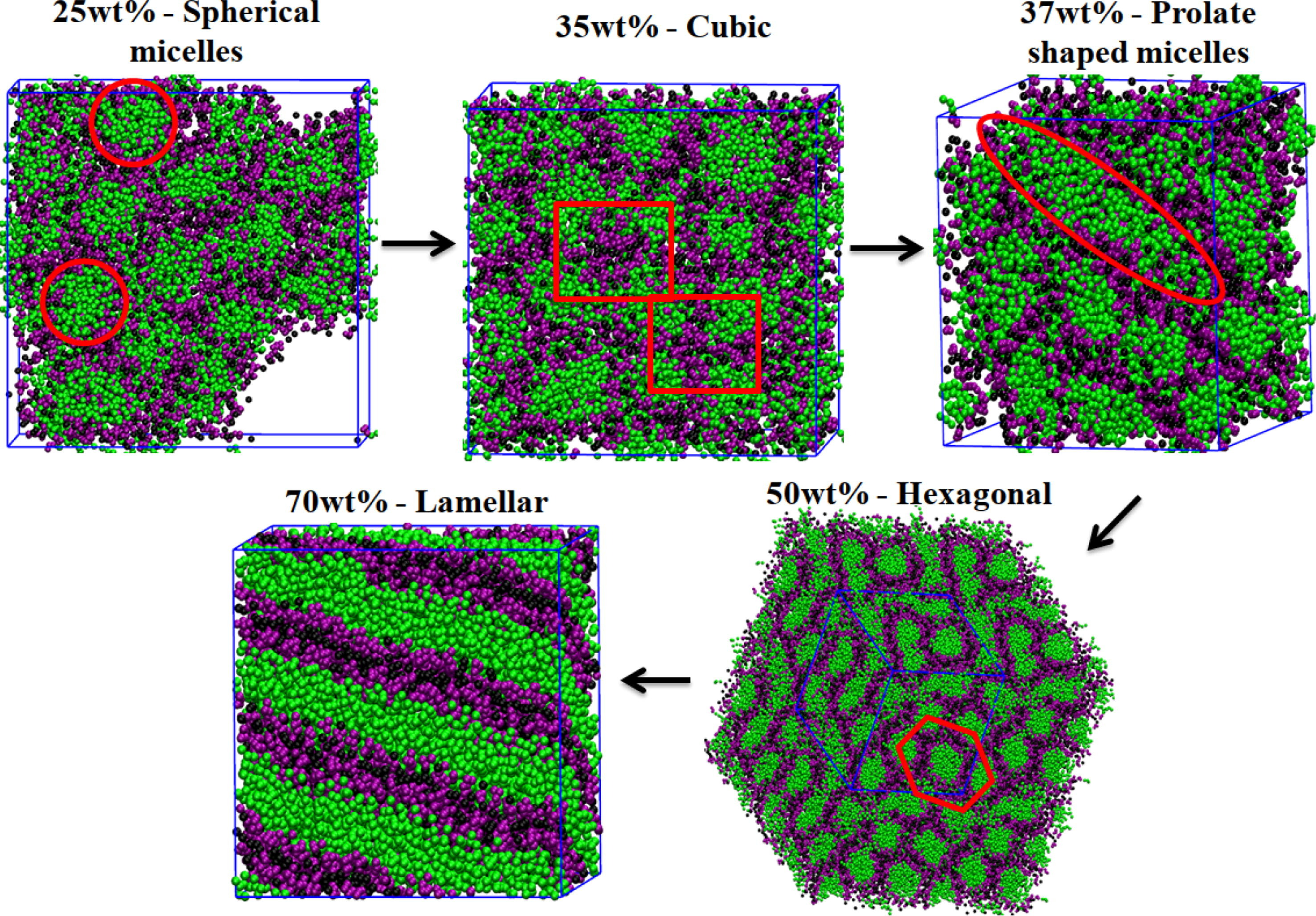}
  \caption{
  Mesophases formed by the ionic liquid ([C10mim][Cl]) water mixtures 
  simulated using Martini for ionic liquids. 
  Reproduced with permission from Ref.~\citenum{2020EACrespo-JColloidInterfaceSci}. 
  Copyright (2020) Elsevier.
  }
  \label{fig:ionic-liquids-mesophase}
\end{figure}

The use of ionic liquids in applications such as extractions is directly linked to the phase behaviour of the ionic liquid as well as to the emerging phase behaviour when combined with co-solvents. Therefore, understanding and predicting these phases is an important step towards efficient computational solvent design. Martini simulations have very recently been used to unravel the phase behaviour of pure ILs, ILs in water, and mixtures of ILs with other molecules.\cite{2020LIVazquez-Salazar-GreenChem, 2019GPerez-Sanchez-JPhysChemC, 2020EACrespo-JColloidInterfaceSci}

For example, \citeauthor{2019GPerez-Sanchez-JPhysChemC} used Martini simulations to understand the temperature dependent effects of adding surface active ILs to Pluronic block-copolymer water mixtures. They were able to elucidate how micelle formation and aggregation changes with Pluronic block-copolymer composition and type of IL. Their results were well in line with experimental cloud point measurements. This example demonstrates that Martini simulations can be used to study temperature dependence of ionic liquid phase behaviour, at least at a qualitative level. Also other studies have successfully used Martini simulations to investigate temperature dependent effects in the context of ILs.~\cite{2020EACrespo-JColloidInterfaceSci, 2019NSchaeffer-PhysChemChemPhys, 2020LIVazquez-Salazar-GreenChem}

For example, \citeauthor{2020EACrespo-JColloidInterfaceSci} investigated the phase diagram of $[C_{n}mim][Cl]$ water mixtures. Their simulations show that these type of ILs display a rich phase behaviour as function of the water content but also temperature in agreement with experimental data where available (Figure~\ref{fig:ionic-liquids-mesophase}). The authors also compare the performance of Martini to bottom-up derived CG models. They found that Martini outperforms those models when it comes to transferability from the neat state to mixtures with water. A similar conclusion was reached by \citeauthor{2020TDPotter-SoftMatter} who investigated the phase behaviour of a type of a non-ionic chromonic molecule. In their study a direct comparison is made between a bottom-up CG model following the approach of \citeauthor{2010LLu-JChemTheoryComput} and the top-down Martini approach to CGing. Only with Martini they were able to simulate the complete phase diagram, whereas the structural CG model showed severe limitations at higher concentrations.~\cite{2020TDPotter-SoftMatter}

Whereas the previous studies focused on ILs in cosolvency with water, \citeauthor{2020LIVazquez-Salazar-GreenChem} used the Martini 3 model to simulate pure $[C_{n}mim[CL]]$. They observed that Martini well reproduces the system density as function of alky chain length and temperature. In addition, the simulations showed the clear formation of dynamic local organization of the IL, so-called nanodomains, which are an important feature of these type of ILs. Furthermore, the authors demonstrated that the Martini model is able to capture a phase transition of $[C_{12}mim[CL]]$~\cite{2020LIVazquez-Salazar-GreenChem}. This phase transition has also been determined experimentally and takes place at around 324.75K. The CG model produced a clear phase transition at 325K, in excellent agreement with experiment.

\begin{figure}[htbp]
  \centering
  \includegraphics[width=0.48\textwidth]{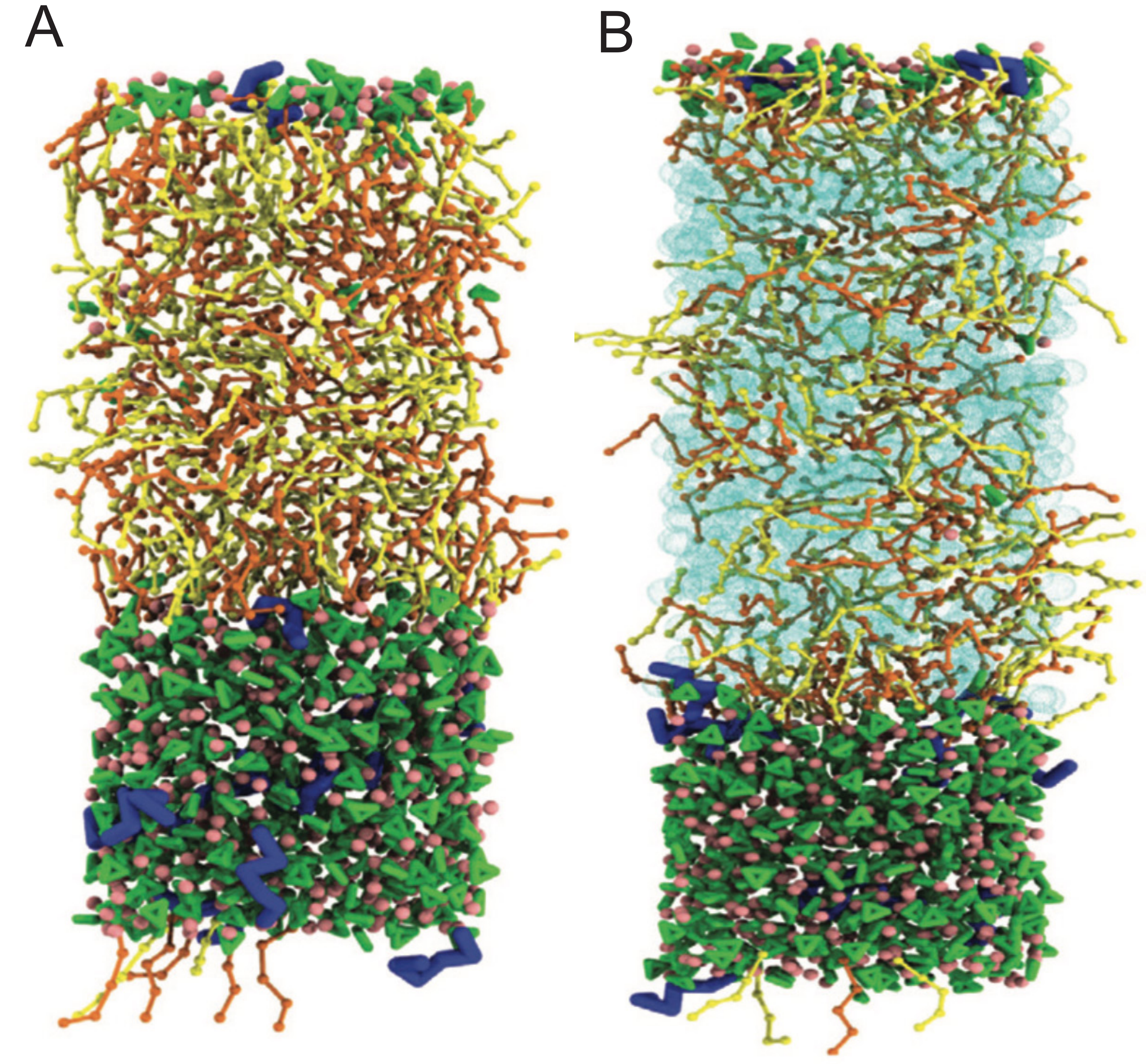}
  \caption{
  Snapshots of the final Martini 3 simulation box of the extraction of poly unsaturated fatty acids from fish oil with an ionic liquids~\cite{2020LIVazquez-Salazar-GreenChem}. In the system (B), octane is added with respect to the composition of system (A), to test the stability of the biphasic system through the addition of co-solvent.
  The color coding is as follows: the IL cation representing the imidazolium ring and the alkyl tail is in green, while the IL anion in pink. The poly unsaturated fatty acids to be extracted are in blue, while the palmitic and oleic acids forming the fish oil phase in yellow and orange, respectively. Octane is depicted in cyan.
   Reproduced with permission from Ref.~\citenum{2020LIVazquez-Salazar-GreenChem}. 
   Copyright (2020) Royal Society of Chemistry.
  }
  \label{fig:ionic-liquids-extraction}
\end{figure}

While the phase behaviour of ILs and their mixtures is an important feature for extractions and applications, it is only indirect evidence for extraction efficiency of a particular IL. \citeauthor{2020LIVazquez-Salazar-GreenChem} also demonstrated that it is feasible to simulate the extraction process directly by creating a biphasic system of the IL and the solvent phase from which the solute is to be extracted (Figure~\ref{fig:ionic-liquids-extraction}). Initially all solute molecules are in the solvent phase, but after 6 microseconds of simulation the solutes distribute between the two phases. From the analysis of the solute and solvent density profiles extraction efficiency and selectivity could be computed. Using this protocol, extraction of benzene and poly unsaturated fatty acids from model oil phases was characterized. It was found that the  simulations, based on the new Martini3 version, are well in line with the trends observed in experiment.

In a different study, \citeauthor{2020GHuet-GreenChem} designed de novo ILs, so called zwitter ionic liquids, which contain the anion and cation in the same molecule. These ILs were synthesized as less toxic and more sustainable variant of $[C_{2}mim][OAc]$. The authors used Martini simulations to assess the toxicity of their de novo designed ILs. It was found that the effect of the new ILs on model yeast membranes was less perturbing than for the original IL. Thus it was concluded that the new solvents are less toxic to microorganisms. These conclusions were verified experimentally by computing the minimum inhibitory concentration.~\cite{2020GHuet-GreenChem} This study illustrates how Martini can be used in a more complete design process to also assess toxicity.
However, ILs are not the only class of green solvents which can be 
simulated with Martini. \citeauthor{2020PAVainikka-DES} have used Martini 
to simulate extraction processes with deep eutectic solvents 
(DES).~\cite{2020PAVainikka-DES} This class of comparatively new molecules 
show similar properties to ILs, however, have advantages in terms of cost 
efficiency and physical properties.\cite{2014SmithChemRev}

\subsection{\normalsize Oils}
\begin{figure}[htbp]
  \centering
  \includegraphics[width=0.46\textwidth]{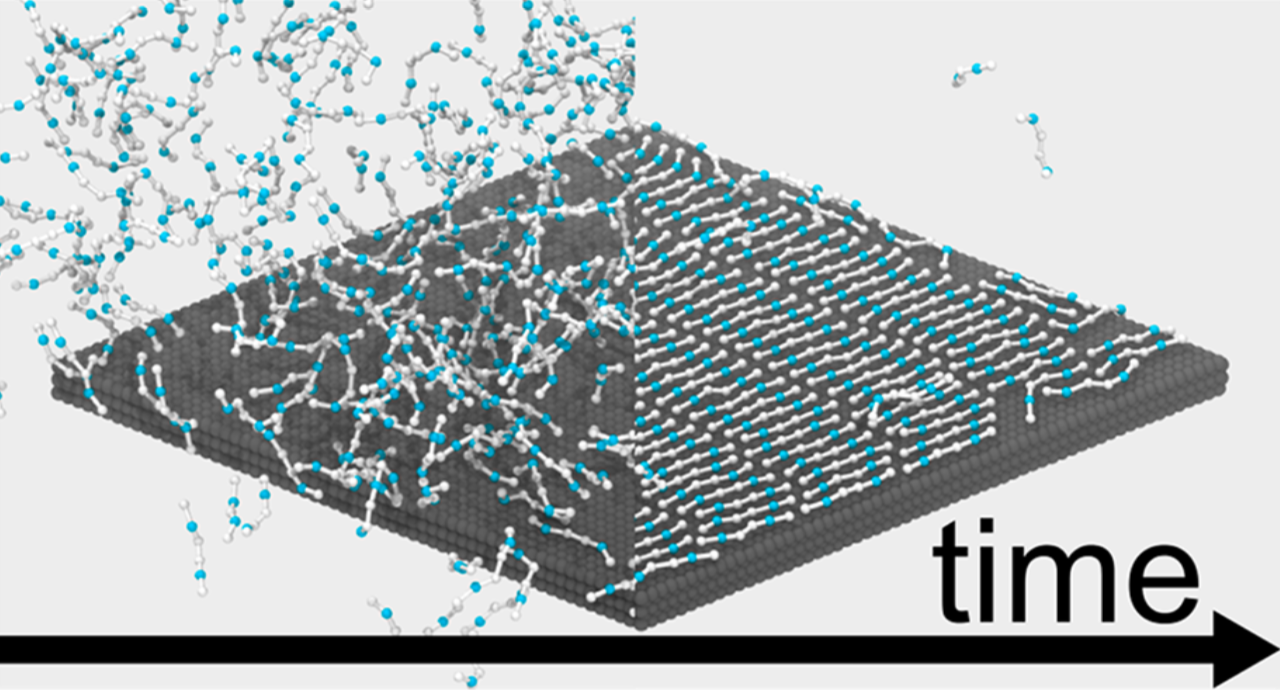}
  \caption{
  Self-assembly on graphite:
  formation of long-range ordered lamellar structures of self-assembling
  (functionalized) alkanes physisorbed on graphite~\cite{2019TKPiskorz-JPhysChemC}. 
  (b) Reproduced with permission from Ref.~\citenum{2019TKPiskorz-JPhysChemC}. 
  Copyright (2019) American Chemical Society.
  }
  \label{fig:alkanes-on-graphite}
\end{figure}

Several Martini-based applications involving oils can also be found in the 
literature~\cite{2016JWang-JPhysChemB, 2017JWang-JPhysChemB, 
2017S-HKim-ChemEngJ, 2018JWang-JPhysChemB, 2018HWang-JPhysChemC,
2020TKlein-JPhysChemB, 2020SKSethi-MolSystDesEng, 2020GLi-ConstrBuildMater,
2020SKSethi-MolSystDesEng}, which are of fundamental interest for 
a variety of industrial processes.
An example are studies of the self-assembly of asphaltenes~\cite{2016JWang-JPhysChemB,
2017JWang-JPhysChemB,2017S-HKim-ChemEngJ,2018JWang-JPhysChemB},
a heavy aromatic fraction of crude oil. 
Their stability in solution strongly depends on temperature, pressure, and 
composition, and is important for the petroleum industry.
\citeauthor{2016JWang-JPhysChemB} found the introduction of partial charges
to represent the radial dipole moment of asphaltenes' aromatic core to be critical
to reproduce the T-shaped stacking behavior observed in atomistic 
simulations~\cite{2016JWang-JPhysChemB}.
Other studies have investigated the self-assembly and self-organization
of various long-chain (functionalized) alkanes 
on graphite (Figure~\ref{fig:alkanes-on-graphite}).~\cite{2013CGobbo-JPhysChemC, 
2018TKPiskorz-JPhysChemC,  2019TKPiskorz-JPhysChemC}
The Martini simulations performed by \citeauthor{2019TKPiskorz-JPhysChemC}
provided a microscopic view on the adsorption and subsequent rearrangement of alkanes on 
the surface to form long-range ordered lamellar structures (Figure~\ref{fig:alkanes-on-graphite}).
The assembly of porphyrin nanorings on graphite has also been 
explored~\cite{2019ASummerfield-NatCommun}.

\section{\large Outlook}
\subsection{\normalsize Martini 3: new opportunities}
Recent identification of some of the limits of the current Martini
version~\cite{2013SJMarrink-ChemSocRev,2019RAlessandri-JChemTheoryComput},
opened the way for the development of a new version, coined
Martini 3~\cite{SOONPCTSouza-Martini3}.
This new version's more general re-parametrization strategy, which did not 
exclusively include biomolecules, is expected to further boost the application 
of Martini in soft materials science.
Areas in materials science which are particularly expected to benefit from the new
re-parametrization are applications involving:
\textit{polymers}, which constitute the backbone of soft materials science
given their high tunability;
\textit{conjugated molecules}, which are ubiquitous in materials given the
possibility of exploiting them as self-assembling systems with interesting
(opto)electronic properties;
and \textit{charged systems}, which are important for applications ranging
from ionic liquids for green solvents to polyelectrolytes for exchange membranes
and next-generation energy storage devices.
Moreover, the Martini 3 parametrization has taken into account not only
infinite dilution properties such as free energies of transfer but also miscibility
data on binary mixtures~\cite{SOONPCTSouza-Martini3}.
As a consequence, the re-calibrated Martini interaction
matrix is expected to perform better in applications involving relative
miscibility, self-assembly, and aggregation propensities.
Additionally, molecular packing is more accurate, as demonstrated
for example in a recent biomolecular study where Martini 3 small molecules were
able to find and bind to protein pockets in a wide range of systems with very
high accuracy~\cite{2020PCTSouza-NatCommun}. Such results are promising
also in view of materials studies were molecular packing is critical: moreover,
the improved molecular packing implies that stacking distances between
aromatic systems, which were off due to the size of Martini small
beads~\cite{2017RAlessandri-JAmChemSoc}, are more accurate in Martini 3.
Overall, we anticipate the new version of the model to show
improved predictions of molecular packing and interactions in general.

In Martini 2, the need for model refinement sometimes led to the development
of custom beads (e.g., see Refs. \citenum{2011GRossi-SoftMatter-PS,
2012LMonticelli-JChemTheoryComput,2018FGrunewald-JPhysChemB,2020TMabuchi-JPolymSci}).
This need often emerged when parametrizing polymers,
where a slight mismatch in the properties of a monomer can
build up into relatively large deviations of the macromolecular properties,
and in some cases due to a suboptimal parametrization of the smaller bead
sizes of the Martini model~\cite{2019RAlessandri-JChemTheoryComput}.
However, the development of custom beads is a time-consuming process and can limit
the applicability of a Martini model if the bead is not validated properly. 
The upcoming new version of Martini 3 includes more generic
interaction modifiers, generally dubbed "labels". Besides the \textit{hydrogen-bonding}
labels already present in Martini 2, which have been expanded and
can now~\cite{SOONPCTSouza-Martini3} be applied to all the new N- and P-bead types,
also \textit{electron polarizability} labels,
which mimic the electron-donor or -acceptor character of certain aromatic fragments,
and \textit{self-interaction} labels, which more generically decrease/increase the
self-interaction of a certain bead type without changing its free energy of transfer,
were introduced~\cite{SOONPCTSouza-Martini3}.
Such labels expand the capabilities of Martini by giving the user a wider selection
of pre-calibrated bead types.
Martini users can now use such extra bead types to fine-tune a certain model.
Accordingly, we expect the introduction of the more generic interaction modifiers in 
Martini 3 to greatly reduce the need for custom beads and allow for quick model refinement.

With the improved balance of interactions and possibility of model refinements,
we expect Martini 3 to be suited to the description of an even wider range of systems.
For example, efforts ongoing in our group are tackling the description of
polyelectrolyte complex coacervates, which have material applications 
in adhesives, coatings, and pharmaceutical applications, 
aedamers, aromatic molecules which mimic biomolecules
as they self-assemble and fold into ordered states,
or metal-organic frameworks (MOFs),
which are of extreme interest for applications such as hydrogen storage
or high-capacity adsorbents for various separation necessities.
Although Martini 3 shows numerous improvements, limitations inherent to this CGing
approach remain. Of general relevance to both material and biomolecular applications
is the limited structural detail due to the CGing process itself. For applications requiring
very fine descriptions, atomistic or structure-based CG approaches are more 
suitable~\cite{2008GAVoth-Coarse-graining-book, 2013WGNoid-JChemPhys}.
Another limitation of general relevance is the entropy-enthalpy compensation.
As the entropy of the system necessarily reduces due to the loss of internal degrees of freedom
upon CGing, it is compensated by enthalpy to reproduce free energies.
Such entropy-enthalpy imbalance is, for example, known to affect the temperature dependence 
of several properties, and should therefore be kept in mind.
More specifically to materials systems, the description of bare metals, such as the one
which may be needed to describe a metallic surface, has not been part of the parametrization 
and, although not impossible, requires careful validation of the chosen bead types.

\subsection{\normalsize  High-throughput materials design}
High-throughput screening of soft matter is an area of immense promise
for materials science.~\cite{2019JJdPablo-NpjComputMater}.
Ideally, a small subset of soft materials would be obtained out of a computational
screening procedure so as to speed up and lower the cost of the experimental step.
Given the versatility and compatibility of Martini and the efficiency gain
with respect to atomistic simulations, Martini simulations are in the position
to contribute to the computational design of soft materials.
Some applications we envision include:
%
\textit{(1) Design of molecular dopants with tailored miscibility:}
molecular doping is an important strategy used to tune organic semiconductor
properties~\cite{2017IEJacobs-AdvMater}.
There are many factors which affect the efficiency of molecular dopants,
one of which is miscibility with the host semiconductor~\cite{2017IEJacobs-AdvMater}.
Martini simulations can be used to screen molecular dopants of different
polarities for insights in their miscibility with a given host semiconductor.
Pushing forward studies such as Refs. \citenum{2017LQiu-JMaterChemA}
and \citenum{2018JLiu-AdvMater}, which investigated the miscibility of only few molecular
dopants, many dopant designs could be inexpensively explored with Martini and
miscibility design rules extracted from such simulations.
Such or similar efforts will have to be coupled with
a parallel screening of said dopants' electronic properties, which could be
obtained by quantum chemical methods.
\textit{(2) Design of green solvents:}
the proof-of-concept showcases of benzene and omega-3 fatty acid extractions
with Martini 3~\cite{2020LIVazquez-Salazar-GreenChem} show promise for the
usefulness of Martini in the computational design of green solvents for
selective extraction using ionic liquids. 
Along the lines of \citeauthor{2020GHuet-GreenChem}~\cite{2020GHuet-GreenChem},
moreover, Martini can be used to design green, bio-compatible solvents
which are less toxic and more sustainable.
Different co-solvents, structural motifs of the ionic liquid, or mixtures could be
computationally screened in order to optimize extraction of certain molecules
or other properties of the solvent such as the predicted toxicity.
Again, we anticipate the recalibrated Martini 3 interaction matrix, validated by taking into
account miscibility data, to be a great asset for such studies.
\textit{(3) Determining molecular structure-morphology relationships in self-assembling materials:}
in order to rationally design self-assembling materials for specific applications,
one needs to derive robust molecular structure-morphology relationships 
of the final aggregate or melt.
However, the chemical space of organic materials is extremely vast, even if one puts
some molecular constraints dictated by the specific application.
Many molecular features are known to affect the final morphology of soft materials,
but extracting robust rules is very time and resource intensive with experiments alone.
Given the compatibility of Martini models and the ease with which one can vary
features such as polarity of molecular features, side chain lengths, molecular topology,
we anticipate that Martini simulations aimed at exploring such parameter space in a
high-throughput fashion are possible.
These efforts, especially if combined with experimental feedback loops and validation,
will help reach the overarching goal of establishing robust structure-morphology relationships
in materials systems ranging from self-assembling supra molecular materials, over thin
films for organic electronics, to nanoparticle-polymer composites.

The above envisioned applications necessarily reflect the own biases of the
authors, and of course many more applications can be conceived along these lines and beyond.

To fully harness the compatibility of Martini models and the growing access to
computational power, and hence realize high-throughput workflows,
the development of tools to automate the Martini workflow will be of key importance.
Such tools need to include programs able to automatically create Martini 
models or Martini building blocks for large molecules. This means they 
need to design the mapping, assign bead types, and derive bonded 
interactions from atomistic simulations. Furthermore tools to generate 
force field files from such building blocks are required as well as tools 
to create initial coordinates for a variety of target systems. 
Such tools would not only accelerate the making of Martini models needed 
for truly high-throughput pipelines, but also make the process 
more robust and reliable.
Endeavors in this direction are already ongoing with tools such as 
AutoMARTINI~\cite{2015TBereau-JChemTheoryComput} or 
Cartographer~\cite{SOONPCKroon-Cartographer}, which are able to derive 
atomistic-to-Martini mappings, as well as perform a basic bead type assignment. 
While the proof of concepts are promising more robust ways for assigning 
bead types and mappings will need to be found. 
Especially with the release of Martini 3, which includes more bead types,
bead sizes, and refined mapping rules.
For parametrization of CG bonded interactions from atomistic simulations
PyCGTOOL~\cite{2017JAGraham-JChemInfModel} 
and Swarm-CG~\cite{2020CEmpereur-Mot-Chemrxiv} have been designed. 
Whereas PyCGTOOL is able to efficiently derive bonded interactions for small molecules, 
it is often impractical for large molecules such as polymers. For example, it cannot 
recognize and optimize redundant bonded interactions. 
This deficit is overcome by Swarm-CG, which uses 
a machine learning-based optimization approach. However, currently generating interactions
for medium-sized molecules is still comparatively slow and not all bonded interaction types 
are implemented. This includes some of those especially important in materials science 
such as the restricted bending potential or the combined bending and torsion 
potential~.\cite{2013MBulacu-JChemTheoryComput} 
Swarm-CG tool at the time of writing is still actively being optimized on this aspect.
Recently, the development of Martinize 2~\cite{SOONPCKroon-Martinize2} aims at mapping 
an entire system from an atomistic reference structure to Martini, generating both target 
coordinates and input files based on already parametrized fragments. 
This will allow to more rigorously transform all-atom systems to Martini resolution 
in one go without having to rebuild the system from the individual components. 
Together with backwards~\cite{2014TAWassenaar-JChemTheoryComput}, this will also allow 
resolution transformation in both directions introducing atomistic detail when needed 
but also capturing that detail when transforming back to Martini.

\begin{figure}[htbp]
  \centering
  \includegraphics[width=0.48\textwidth]{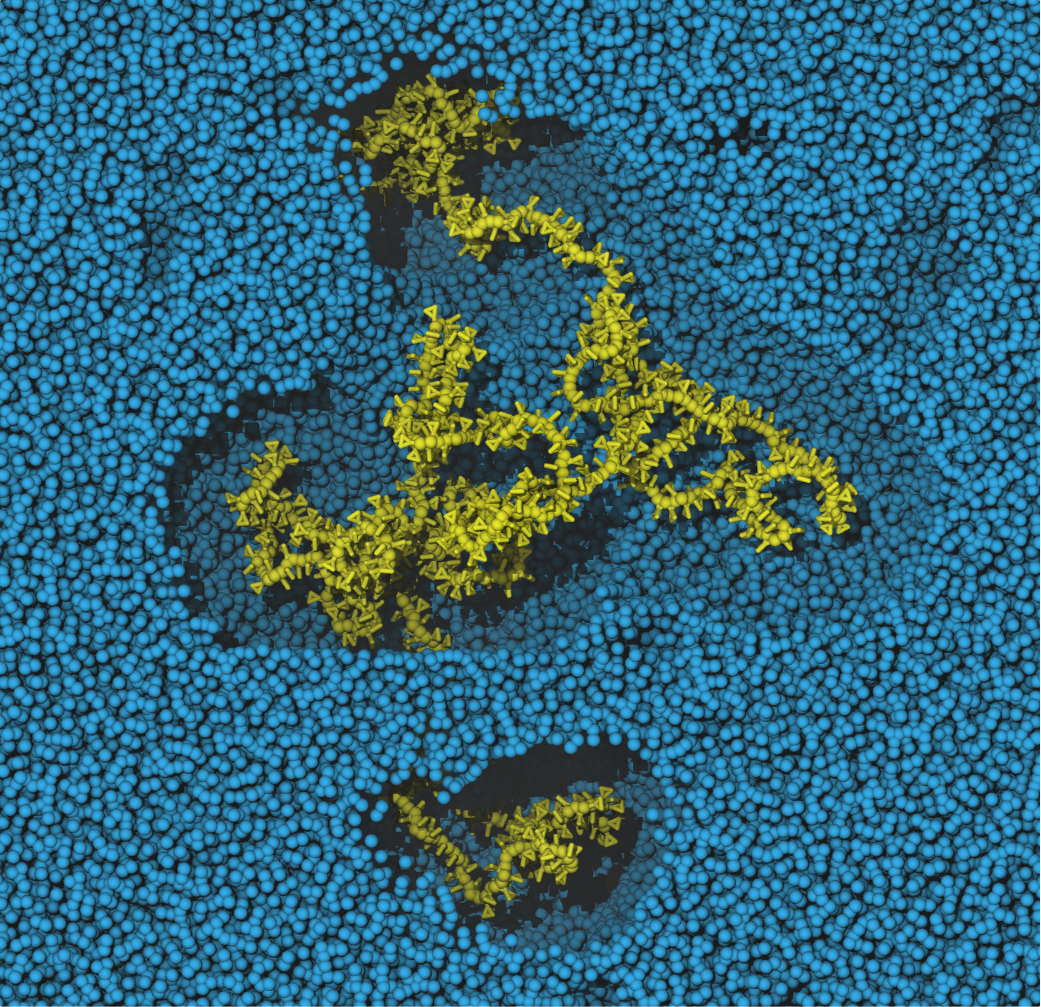}
  \caption{
  Melt of Martini $PS_{1000}$ generated using Polyply~\cite{SOONFGrunewald-Polyply}. 
  Highlighted in gold is a single chain with other chains (blue spheres)
  removed within a cut-off of 2.3 nm. 
  The total system size is about (30 nm)$^{3}$.
  }
  \label{fig:Polyply}
\end{figure}

Besides model parametrization and resolution transformation, tools to setup starting 
structures will also be of increasing importance as the complexity of the simulated systems grows. 
In this context, recently \citeauthor{SOONFGrunewald-Polyply} developed Polyply,
a software suite for facilitating atomistic and CG polymer simulations.  
The tool can generate topologies of polymer systems ranging from simple or complex homopolymers, 
over branched and hyper-branched polymers, to block copolymers~\cite{SOONFGrunewald-Polyply}. 
Not only single chain starting structure but also melts and more complex pre-assembled 
morphologies can be generated. The latter strategy is useful, 
because whereas self-assembly might be the preferred strategy to assemble a polymer morphology, 
it remains a challenging task even at the Martini CG level due to the slow dynamics 
of long polymer chains. 
For example, Figure~\ref{fig:Polyply} shows a PS melt system with a molecular 
weight of 1000 residues. Highlighted in gold is a single chain, which winds and twists 
almost from one edge to the other. The total dimensions of the system is about 
(30 nm)$^{3}$ and comprises half a million CG particles. 
While properly mixing such a melt from initially disentangled chains 
is an almost impossible undertaking, Polyply generates such structures within minutes. 
Polyply's internal library already contains several Martini polymer models from the literature 
but more models can be contributed via GitHub~\cite{SOONFGrunewald-Polyply}. This tool 
is expected to standardize and greatly simplify 
the generation and setup of Martini polymer systems.

Other tools with a similar purpose are also being developed. For example, CHARMM-GUI already
supports generation of Martini membranes~\cite{2015QiJCC} and sugars~\cite{2017HsuJCC}, 
and is currently being extended to polymers. Whereas Polyply and CHARMM-GUI can generate 
both topologies and structures, PACKMOL~\cite{2009MartinezJCC} is a tool which can be used to
generate starting structures from existing molecule coordinates.

Maintenance of a library of available Martini models, and their curation, 
will also be important on this front. Many Martini models are available 
at \href{http://cgmartini.nl}{http://cgmartini.nl}: besides an extensive
number of biomolecules, and in particular lipids, a growing ``polymerdome'', and upcoming 
extensive Martini 3 solvent~\cite{SOONPCTSouza-Martini3} and 
small-molecule~\cite{2019RAlessandri-PhDthesis} databases 
will be important for materials science applications.
Moreover, both the upcoming Martinize 2~\cite{SOONPCKroon-Martinize2} and
Polyply~\cite{SOONFGrunewald-Polyply, 2021FGrunewald-chapter} tools 
rely on the Vermouth library~\cite{SOONPCKroon-Martinize2}, 
which offers a consistent way of defining building blocks and keeping track of model versions, 
thereby contributing to better data curation which is an important aspect of growing data sets.

\subsection{\normalsize  Advanced Martini simulations}
The growing computational power available and the increase in the
complexity of the simulated systems demands for smart and (semi-)automated ways to drive,
explore, and analyze the simulated system. Moreover, combining Martini with new method
developments can open the way for advanced simulations which go beyond
what is feasible with standard CG MD.

Despite the growing computational power available, brute-force sampling
of the conformational space often is still not sufficient.
Enhanced sampling techniques are available to increase the effective
simulation time of MD simulations in general. A host of enhanced sampling techniques already exist
and it is in continuous expansion, and many techniques are implemented in packages such
as PLUMED~\cite{2014GATribello-ComputPhysCommun}
and SSAGES~\cite{2018HSidky-JChemPhys}, which are
compatible with many MD softwares, including GROMACS~\cite{2015MJAbraham-GMX5}
and NAMD~\cite{2005JCPhillips-NAMD},
and hence can be readily applied to Martini systems.
Given the insights such techniques have already provided when applied
to Martini simulations~\cite{2017DBochicchio-NatCommun, 2018SHJung-JAmChemSoc}
and the active developments in this area,
Martini simulations will surely benefit from coupling to such techniques.

\begin{figure}[htbp]
  \centering
  \includegraphics[width=0.48\textwidth]{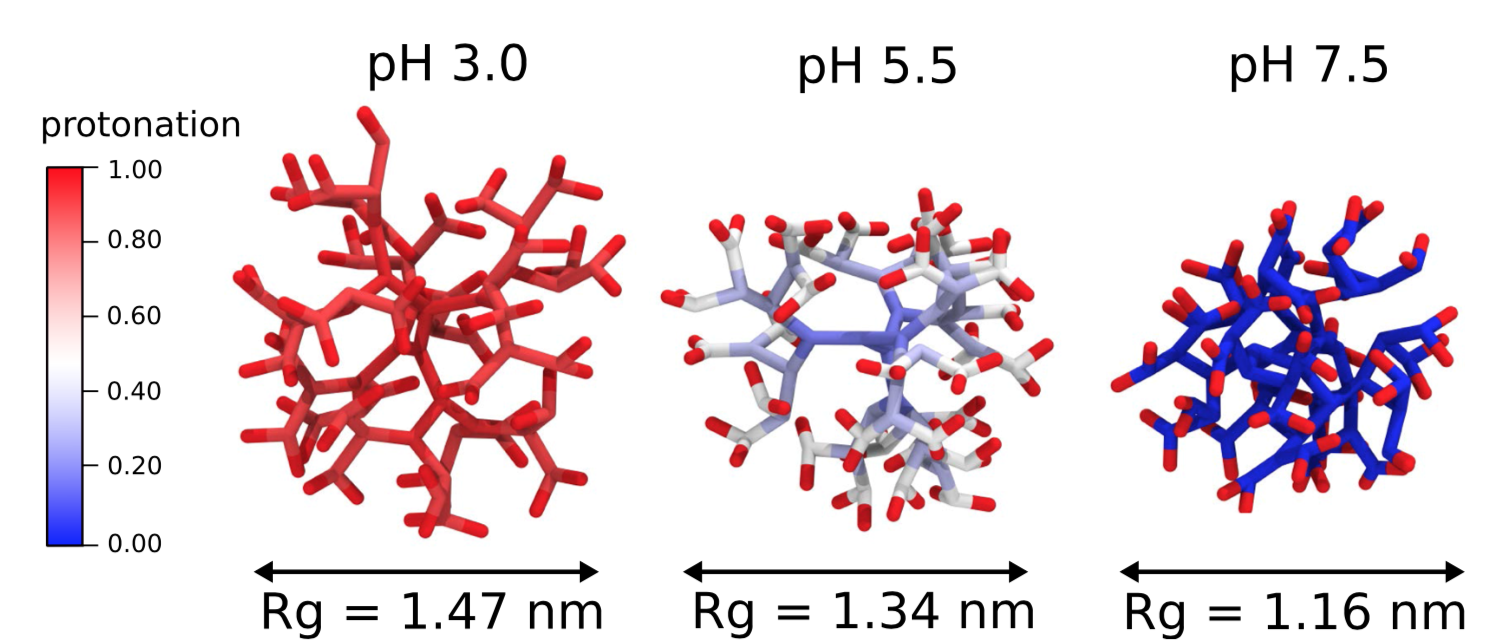}
  \caption{
  Protonation of a poly(propylene imine) (PPI) dendrimer as a function of pH as 
  modelled with the titratable version of Martini~\citenum{2020FGrunewald-JChemPhys}.
  The protonation state of the core beads of the dendrimer, which represent tertiary amines, 
  clearly change as function of pH becoming progressively less protonated as indicated with 
  the color scale from red (protonated most of the time) to blue (deprotonated most of the time). 
  The radius of gyration (Rg) quantifies the degree of polymer collapse as the charge 
  density decreases at higher pH. 
  Reproduced with permission from Ref.~\citenum{2020FGrunewald-JChemPhys}. 
  Copyright (2020) American Institute of Physics.
  }
  \label{fig:Martini-titratable-dendrimer}
\end{figure}

Changes in bonded interactions and atom types, which reflect chemical reactions, 
cannot be captured by regular MD simulations.
However, chemical reactions especially in the field of material science are ubiquitous. 
Examples include cross-linking reactions in polymers, dynamic changes in the protonation 
states of polyelectrolytes, or formation of gels. 
Efforts are being taken towards capturing these effects.~\cite{2011GRossi-Macromolecules, 
Zadok2018, 2019HGhermezcheshme-PCCP, 2020FGrunewald-JChemPhys} 
\citeauthor{2011GRossi-Macromolecules} studied the cross-linking of a polyester resin. 
To model a cross-linking reaction they used an ad-hoc empirical potential, which displayed
harmonic-like features at close distance, but was able to dissociate at larger distance. 
In contrast, \citeauthor{Zadok2018} used a LAMMPS built-in feature to simulate the effects of 
gel formation for a protein imprinted hydrogel.
\citeauthor{2019HGhermezcheshme-PCCP} used a similar approach, however, combining 
GROMACS with an in-house code to study the step-growth reaction of polyurethane. 
Here a bond is introduced after finding all reactive neighbors within a cut-off. 
Upon changing the topology a long equilibrium simulation is carried out 
after which another step of cross-linking is performed. 
In contrast, the recently developed titratable Martini~\cite{2020FGrunewald-JChemPhys} allows the
dynamic representation of protonation reactions in a Martini simulations. 
Here the protonation state of a titratable functional group can change back and forth between 
protonated and deprotonated during the course of a continuous simulation. Using this approach for
example, the pH induced collapse of the hyper-branched polymer poly(propylene imine) 
was simulated as shown in Figure~\ref{fig:Martini-titratable-dendrimer}. 
Like the approach by Rossi an empirical nonbonded potential is used to allow protons 
to tightly bind to titratable functional groups. Further progress in the field of
reaction simulations is expected with the inclusion of lambda dynamics into 
GROMACS~\cite{2015MJAbraham-GMX5}.

Machine learning (ML) approaches coupled to molecular modeling are emerging
in soft materials research~\cite{2018ALFerguson-JPhysCondensMatter,
2019NEJackson-CurrOpinChemEng}.
Such approaches have the potential to support and augment traditional physics-based models
in computational research~\cite{2018KTButler-Nature}. 
In the realm of soft materials, ML techniques have been shown to
be able to predict electronic properties, such as energy levels and absorption spectra,
of soft materials directly from CG structures~\cite{2019NEJackson-SciAdv, 2020LSimine-PNAS}.
This strategy is a potentially quicker and less laborious alternative to the currently
necessary backmapping and subsequent quantum chemical calculation step,
and it is especially relevant to systems where Martini structures are usually
backmapped to obtain electronic properties, such as organic electronic
systems~\cite{2015MBockmann-PhysChemChemPhys, 
2020NRolland-ComputMaterSci, 2020RAlessandri-AdvFunctMater}.
Another interesting avenue is the one of coupling CG simulations and
ML techniques to efficiently explore the desired chemical space.
For example, Shmilovic and co-workers~\cite{2020KShmilovich-JPhysChemB}
used Martini simulations within an active learning strategy to efficiently
cover the chemical space span by $\pi$-conjugated peptides with
a $\pi$-core flanked by two tripeptide units.
Instead of simulating all possible tripeptide sequences ($20^3 = 8000$),
the active learning strategy allowed to minimize the number of simulation data
needed for training the ML model. Accordingly, by direct simulations of only 2.3\%
of the tripeptide space, the authors showed how a Gaussian process regression
model could capture the statistical information necessary to represent
this chemical space~\cite{2020KShmilovich-JPhysChemB}.
The model could then be leveraged to identify the $\pi$-conjugated peptides predicted
to exhibit superior assembly properties to those reported in previous work.
Moreover, in this way the authors were able to reveal design rules governing assembly
of these molecules.
The fact that the Martini bead types discretize the chemical space,
means that similar molecules will often map to the same Martini CG model.
This introduces a degeneracy in the CG representation, which translates into
a reduction of the size of the chemical (compound) space.~\cite{2020TBereau-arxiv}
Accordingly, this reduction of the size of the chemical space represents 
also a further speed-up which can help for screening studies.~\cite{2020TBereau-arxiv}
Hence, we expect hybrid Martini/machine-learning schemes to be highly
promising in order to efficiently explore the chemical space for different applications.

\vspace{5mm}

In conclusion, we foresee a bright perspective for applications of the Martini model
in the field of materials science, especially given the possibilities offered
by the upcoming new version of the force field, existing and forthcoming tools
to streamline model building and system preparation, and
combination with existing and future method developments, such as constant
pH simulations and hybrid simulation/machine learning schemes.

\newpage
\providecommand{\latin}[1]{#1}
\providecommand*\mcitethebibliography{\thebibliography}
\csname @ifundefined\endcsname{endmcitethebibliography}
  {\let\endmcitethebibliography\endthebibliography}{}

\end{document}